# Evaluation of isolation design flow (IDF) for Single Chip Cryptography (SCC) application

By

Arsalan Ali Malik

UET-17S-MCE-CASE-007

Supervisor

Dr. Anees Ullah

DEPARTMENT OF ELECTRICAL & COMPUTER ENGINEERING

CENTER FOR ADVANCED STUDIES IN ENGINEERING (CASE)

UNIVERSITY OF ENGINEERING AND TECHNOLOGY TAXILA PAKISTAN

FALL 2019

# ABSTRACT

Field Programmable Gate Arrays (FPGAs) are increasingly in various applications. This is due to the fact that they provide flexibility to reprogram and modify in real-time with minimum effort. The increasing usage of FPGA must also ensure its end users with a guarantee that they are able to overcome failures and work in harshest of environments. Now-a-days end users demands a guarantee that the system he/she is buying remains functional.

FPGA Vendor Xilinx provide its users with that guarantee in the name of Isolation Design flow or IDF for short. Xilinx claims that IDF can help ensure users the reliability all the while providing its flexibility. If true, application that can benefit from IDF are vast such as Avionics, Unmanned probes, Self-driving fully autonomous vehicles etc.

This thesis puts the Xilinx claim regarding IDF to test by implementing a single chip cryptographic application namely Advanced Encryption Standard according to the rules and regulations defined by isolation design flow. This thesis does so by replicating and injecting faults in the system that conforms to the rules of IDF and records its behavior first hand to observe IDF effectiveness. This thesis can also help end users and system evaluators interested in effectiveness of IDF with an independent view other than Xilinx by providing them with a statistical data collected to remove all doubts.

## ACKNOWLEDGEMENTS

First I'd like to thank my graduate advisor Dr. Anees Ullah for the vision and direction that helped this thesis become a great triumph. There were various instances that I had doubts or questions about the path this work should take; however, a quick discussion with him always helped me overcome these hindrances. I'm also grateful for the many hours he put into refining and refocusing my writing despite his relentlessly busy schedule.

Secondly, I'd also like to thank my family and friends for their endless support and patience throughout this long process. They motivated me to put in the long hours necessary to make this thesis a reality. This accomplishment is as much their as it is mine.

# Table Of Contents

# List Of Figures

# List of Tables

# CHAPTER 1 : INTRODUCTION

FPGA have revolutionized the field of embedded system by providing the flexibility of reconfiguration in real-time. FPGA’s are widely used nowadays in different precision critical fields such as avionics, un-manned probes, and cryptographic environments. FPGA has allowed to merge multiple chip solution onto a single chip. Multiple chip solution that were once area and time hungry have shrunk onto a single tiny chip.

FPGA’s are considered as a cost-effective replacement for ASIC’s as they provide ease for user on two fronts. First they provide users with the flexibility to replace or amend their logic in field reducing lengthy time involved in the process of ASIC manufacturing and development. Secondly cost incurred in procurement of FPGA’s is much less then developing ASIC from the scratch. These features of FPGA also poses a challenge in areas such as safety critical application and government/military sectors where a need for a reliable system is dire.

As an end-user utilizing the FPGA, the demands are forever increasing. End user demands a guarantee that the system he is buying remains operational no matter what as he can neither tolerate nor afford failure in field. Multiple chip solution which provided design isolation and fault containment inherently in the case of system failure are being replaced by a single chip solution quickly. Hence, the requirement of modular level design isolation in a single chip solution, harboring multiple functionalities has risen.

Over time various solutions for avoiding a single point of failure in a designs have been proposed and adapted. Xilinx provides its user this in the form of Isolation Design flow (IDF). Xilinx claims that IDF can provide the design segregation, isolation and reliability in harshest of environments. If true, application that can benefit from IDF are vast such as Single-Chip Cryptography, Avionics, Modern Warfare, Artificial Intelligence etc.

## 1.1 Motivation

Xilinx has marketed major uses of IDF repeatedly. The problem with this advertisement is, besides Xilinx we did not find any data that supports this claim thus we decided to evaluate the effectiveness of Isolation Design flow first hand to provide system evaluators and end-users with an unbiased third-party opinion. This research is focusing on Xilinx, one of the leading name in FPGA chip fabrication and designer giant in industry. This research prime focus is to test and evaluate one of the reliable system design techniques introduced by Xilinx named; "Isolation Design Flow" (IDF) and its practical aspect for single chip cryptographic application. Purpose of the research is too verify first hand, that whether or not all the statements that are claimed by this leading manufacturer hold true or not.

IDF is a technique which allows users to postulate, identify and isolate core logic in a system which if would have left unchecked; can result or in leaking of un-intentional information to unauthorized user or in some extreme cases, complete system breakdown/meltdown. As per Xilinx, IDF is pioneered to be used in areas where security is of upmost importance to the user such as government sectors. IDF has opened a doorway to designing a secure single chip solution in which designers can combine multiple functionalities, each isolated from one another. Isolation between various function is ensured with the help of *fence(s)*. Fence are set of tiles that are marked unusable by user in designing stage by careful *floor planning* of design, restricting the routing or placement of user logic from said tiles.

Should a communication between isolated modules is required, it is done so by the use of trusted routing. IDF advises users to minimize the placement of logic on top level as much as possible. In fact Clock signals are the only ones that are recommended for placement in top level logic. Xilinx imposes several rule that are to be followed for using IDF in a design. But what if after following all these rules there exist a possibility that your design can be hacked, leak information or break down in the occurrence of single event upset (SEU). What is the success rate of IDF? How effective is IDF against fault injection methods? What events could lead to drift open the isolation between modules? These are the questions that are not answered by Xilinx and we aim to find out by fashioning a unique blend of IDF and partial reconfiguration (PR). We venture to introduce faults in an isolated design created using IDF and introduce a novel approach to recover from complete system breakdown in the event of IDF failure. We are testing the limits of IDF all the while providing a solution to enhance the capabilities of IDF with the help of PR.

The primary research work that this studies entails is to develop an understanding of how isolation design flow works and test its capabilities. The complete configuration of the FPGA will be put under various fault injections to test and trial underlying IDF effectiveness. Result of these attacks on FPGA's running configuration will be observed by performing real-time comparison with the golden bit stream mentioned in previous steps.

## 1.2 Objective

Xilinx has proposed a standardized solution for its users to eliminate the need of creating a different fail-safe solution every time system reliability is to be achieved. Xilinx's solution to their problems is called isolation design flow (IDF). Xilinx claims that by presenting IDF it has introduced a single chip information assurance solution which was not attainable previously and users were forced to achieve this feature by using multi-chip solution or integration of distributed solution.

This research sheds light on one of the technique developed to provide users with peace of mind and reliability in the field named Isolation Design Flow and evaluated its performance by putting it to test under normal and duress conditions to check its reliability.

*Does the IDF technique developed by Xilinx is really providing design level fault tolerance?*

Xilinx claims to have achieved fault tolerance by introducing IDF as a solution. As per Xilinx; IDF can provide users fault containment capability in FPGA at modular level by incorporating techniques such as:

(a) Modular level redundancy
(b) Watch-dog timers/alarms
(c) Seclusion by safety levels
(d) Isolation of test logic (i.e. for harmless/safe removal)

These techniques, if properly utilized can help provide FPGA designer satisfaction, providing him an environment which is fault tolerant enabling him to focus on other challenges involved in system designing process.

*Solution Methodology:* To test the extend of IDF reliability we have developed an interface which will acts as a door way to user logic and contaminating FPGA logic and observe it ability to recover from the faults introduced.

*Is IDF prone to single point of failure (SPOF)?*

Before the introduction of IDF, the reliability in a system and elimination of SPOF was primarily achieved by using techniques such as double modular redundancy or triple modular redundancy (TMR), scrubbing, device level fault tolerance, self-repairing/healing architecture etc. To evaluate IDF against such techniques, different attacks and fault injections is mandatory.

*Solution Methodology:* An interface such as PCAP or ICAP gives access to FPGA fabric once the design is configured. We will be injecting faults using PCAP and observe the IDF and its ability to cope up against faults introduced.

We are basing our research on the application note provided by Xilinx as a reference for isolation design flow. We assume that the design provided by Xilinx is error free and conforms to all the rules and regulations necessary for IDF. By using the reference application note provided by Xilinx we are also ensuring that we are able to compare our design conform to the IDF’s rules as implemented in the reference design provided by Xilinx.

## 1.3 Thesis Organization

The thesis is designed in the following sequence; Chapter-1 provides the introduction and promotes the need of the research, Chapter-2 covers the literature review and background research carried out to complete this topic, Chapter-3 discusses the test setup and research method used to carried out the research with Chapter-4 providing detailed analysis on the results conducted. Finally, Chapter-5 concludes the results of this findings along with future work to be done in this field.

# CHAPTER 2 : LITERATURE REVIEW

FPGA's were designed to remove the limitations of ASIC's by providing the flexibility to reprogram and reconfigure in run-time; as per user needs. With the emergence of cloud computing and IoT's FPGA's are now widely used as hardware accelerators to off-load intensive computational tasks to FPGA's reducing time and saving resources.

BBC; one of the radio channel pioneer is also utilizing Zynq SoC to leverage FMC for audio codec replacement, switching 32-years old rack-mount audio codec with commercial of the shelf component. Example of such applications is *Enyx*, a company which provide users with an online FPGA hardware acceleration services by providing them with software/hardware experience of their own choosing. User can just drag and drop the type of service desired and it will be provided in shape of Intellectual property (IP's) tailored specific to their needs.

Now-a-days with the invention of partial reconfiguration (PR), many user share same FPGA chip carrying out their desirable operations side by side. PR has opened doors to an entire new threat models, originally not present in traditional FPGA's. Zao et-al work shows that side channel attack on such environment is easily possible by using ring oscillators (RO), placing them in close proximity to secondary user's space with the sole purpose of snooping data and information gathering/leaking [1].

Their RO based design acts as a power monitoring device which can relay information such as switching of bits from 0 to 1 which can be used to make a dictionary model and co-relation can be applied on it to extract the actual data. Moreover such design can be placed without using place and route constraints. Hence isolation between users to prevent such attacks is necessary.

We hope to evaluate such isolated design created by Xilinx IDF to help users around the globe secure their design against threats such as side channel analysis, power monitoring etc.

## 2.1 System Design Challenges

From the literature review we have gathered the following considerations that must be taken care of when developing a reliable, field capable, isolated system. When creating an FPGA or SoC based design, the fundamental things that are on top of the designer's head are area, time and power. These constraints are what limits the broader view of a designer. However in order to create an isolated, robust and more long term, many aspects must be considered that are captured in figure below and explained in subsequent sections.

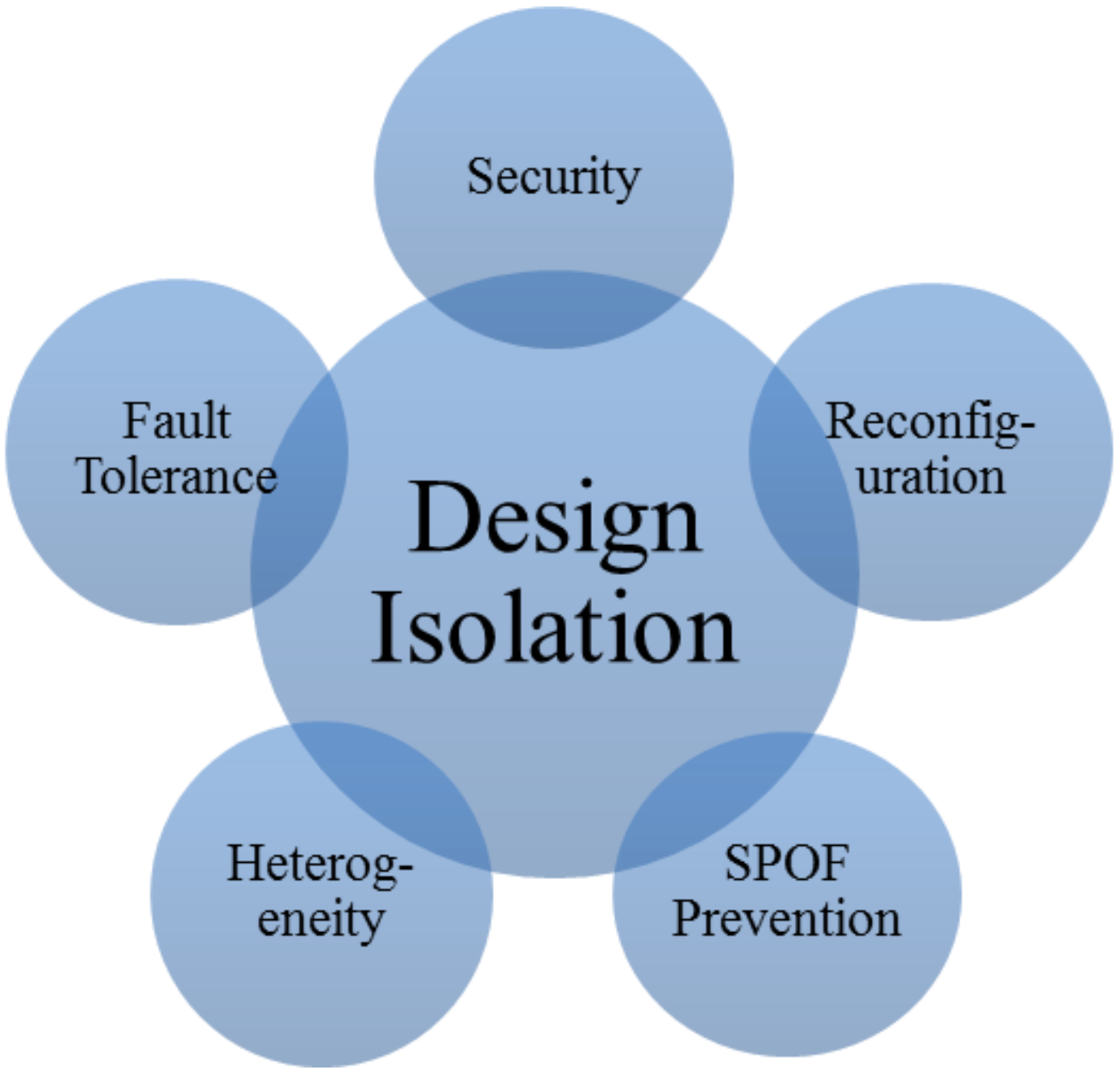

Figure 2-1 : Challenges in Reliable System Designing

### 2.1.1 Security

Design security is a critical need of many industries and classified military applications. To this purpose encryption is being used to mask actual data from the adversary. However, with the emergence of side channel analysis, researchers have proposed and evaluated several designs, exposing their weaknesses. Rubén Lumbiarres et al [3] has created an assistive method for such purpose termed as "Faking countermeasure". Their proposed solution is processing the unencrypted or plain-text data with the help of a false/fake key whose EM wave emission will mislead the attacker. The False key used for the purpose holds the $K_{FALSE} = K_{REAL} \oplus K_{MASK}$ relationship with the original key.

The additional operation needed at the end of each mix-column operation is exclusive-or of the false output with $K_{REAL.}$ This processed adds a huge overhead for large chunks of data is primary focused on the AES design thus cannot be extended to other ciphers schemes. In the past, NSA has worked with Xilinx to secure Virtex-4 against attacks mentioned in special cover feature paper *FPGA Security: Motivations, Features, and Applications* [29]. The work performed was focused on evaluating isolation of basic building block of an FPGA; CLB's and Global switch matrix (GSM) which is used to facilitate interconnect between CLB to CLB. The idea was *Fence* to provide physical isolation was introduced in this paper. Xilinx has further improved on this idea to create IDF [13] , [22].

### 2.1.2 Reconfiguration

Reconfiguration of FPGA is arguably "*The Selling point"*. The flexibility provided by FPGA is what attracts the vast amount of users towards it. Reconfiguration has enabled FPGA to gain a massive acceptance among big data analytics, parallel and distributed computational systems. FPGA now-a-days hold multi-tenants in cloud computing as their general purpose computation machine. PR is one of the leading name in the cloud computing. PR allows user to time multiplex FPGA resources enabling effective use of chip logic density. The reconfiguration time has thus reduced as the partial bit stream size is fairly smaller than the whole bit stream [4]. PR allows designed to be portioned in *Static* and *Dynamic* parts.

The dynamic or partial reconfigurable module can be arranged on chip in various configurations (As shown in Figure 2-2).[6] Each style has its own pro's and con's. Island style is the most simplest to implement but suffers from internal fragmentation resulting in high percentage of resource wastage. Slot style configuration does not have fragmentation problems where module can occupy resources as per its need. Tiling of RM region is a very complex task in which one has to keep in mind the placement of routes and their cross-over from static to dynamic regions. The optimal

PR style to use depends on the user requirement and may vary or consist of a model that is hybrid of two or more styles.

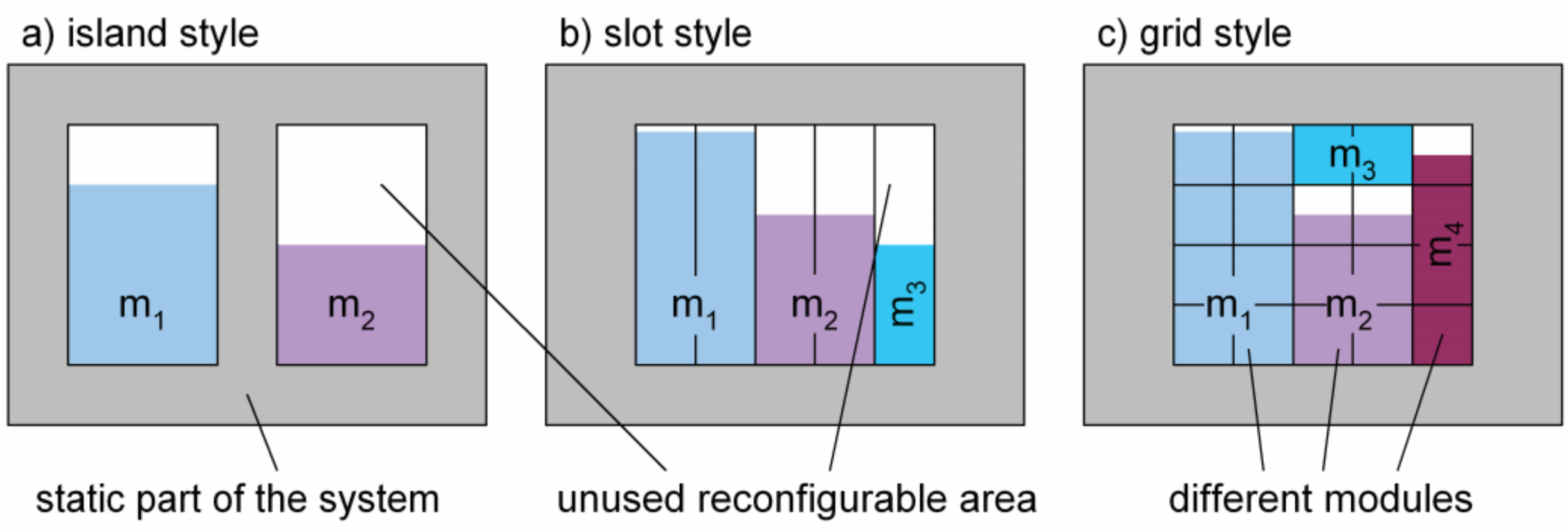

Figure 2-2 : Reconfiguration Styles [4]

### 2.1.3 Single point of Failure (SPOF) Prevention

Embedded systems are designed and deployed every day to be used in every walk of life ranging from sub-marines to a handheld smart phone. If proper isolation and fault control mechanisms are not kept in mind during the time of designing, the damages could be catastrophic. SPOF are unwanted in any system, systems are designed with the ultimate goal of great reliability. To avoid SPOF in systems various solutions have been provided from time to time. SPOF preventions can be put in place at (a) Internal components level (b) System level (Distributed Systems) (c) or a site wide level (Repetition).

In the work provided in [23] SPOF prevention method is introduced by building upon the existing TMR methods. In this research, author has pointed out the slow configuration rate provided by commercial available tools due to using only majority voters. The author proposes the use of minority voters along with the majority voters to detect fault at granular level of FPGA. The technique can detect and help user correct the fault by using the output of minority voters as a *flag*. Should the system fail to correct the SPOF it can minimize the damages if only the systems is implemented in isolation preferably at both physical and logical level.

### 2.1.4 Heterogeneity

Earliest FPGA developed were homogenous devices. Now-a-days various functionalities are bundled down on one single chip. Isolated design helps achieve the best of FPGA homogeneous nature where user can access and explore DSP, BRAM, and CLB all in one place. In heterogeneous system on chip (HRSoC) device, user application can be easily dived into various tasks which can be either hardware accelerators based which are built-into device at pre-defined partition or location or can be implemented as a software task based on computing elements or CLB's.

Gantel et al in his paper [25] has made use of heterogeneous nature of FPGA along with IDF to combine flexibility and reliability. Gantel used two isolated partitions (separately) that are identical to each other (resource wise) to ensure that relocation process is achievable without physically damaging of chip. Just like in IDF trusted routes are established to communicate various modules of a partition. A signal that is sink to two or more module is split into different signal(s) and passed through LUT resources to form trusted routes.

Use of hard macros such as *INST <Hard Macro> LOC =* $SLICE_X^{a}Y^{a}$ where "*a*" is the valid XY coordinates of chip is done to constraint synthesizer for correct desirable position. The Isolation of designs ensured the error free relocation of modules in accessible dynamic partition space.

### 2.1.5 Fault Tolerance

The ability to control system failures is a desired feature that has promising aspects in government cryptographic systems, avionics and functional safety electronics. Xilinx claims that by using Isolation Design Flow [24] one can implement fail-safe in user logic to avoid total system failure and enhance system dependability.

If a user can make system failure(s) reliant on numerous autonomous subsystem failures, trustworthiness and dependency can be enhanced by many levels of degree.

Example for this problem is beautifully illustrated in [32] , *a system which is composed of two or more redundant subsystems connected in parallel fashion has a* *failure probability which is equals to the multiplicative product of the probabilities of each of the subsystems failing*. i.e. If the subsystems both had failure probability of $10^{-9}$, then the system made up of these subsystems has a reduced failure probability of $10^{-9} * 10^{-9} = 10^{-18}$, which is many fold lower in magnitude than each subsystem failure rate [32].

However, this calculation hypothesize that probabilities of failure are polar to each other in nature i.e. the subsystems does not have a SPOF or common failure mode. This notion is the genesis of isolation design flow.

A fault tolerant system design must keep in mind the following considerations [32]:-

**Security:** Authentication, confidentiality, integrity, non-repudiation, availability and utility.

**Fault-tolerance:** Failure mode and effect analysis, configuration scrubbing, floor-planning, module decomposition, reduced functionality modes, built-in self-tests (BIT) fault containment, failover/failback, redundant alarms, configuration memory error detection and error-recovery /correction mechanism.

**Validation and verification:** Automated test procedures, diagnostic logging, design for test, and formal verification

## 2.2 Fault Injection Techniques

To test and evaluate the IDF we decided to inject fault in the PL side while recording its response from PS side creating a statistical database which can help readers understand the necessity of design isolation and the effectiveness of Xilinx IDF solution. By literature reviews we found that there are several methods out of which following five methods were explored which are as follows:-

### 2.3.1 JTAG

The Joint-Test Action Group (JTAG) interface is a well-known, easy-to-use, serial interface that provisions the IEEE 1149.1 standard for boundary-scan and Test Access Port architectures. JTAG is a most widely used industry standard that is used for design verification and testing printed circuit boards after they have been manufactured. JTAG encompassed standards for on-chip instrumentation in electronic design automation (EDA) as a complementary tool to digital simulation.

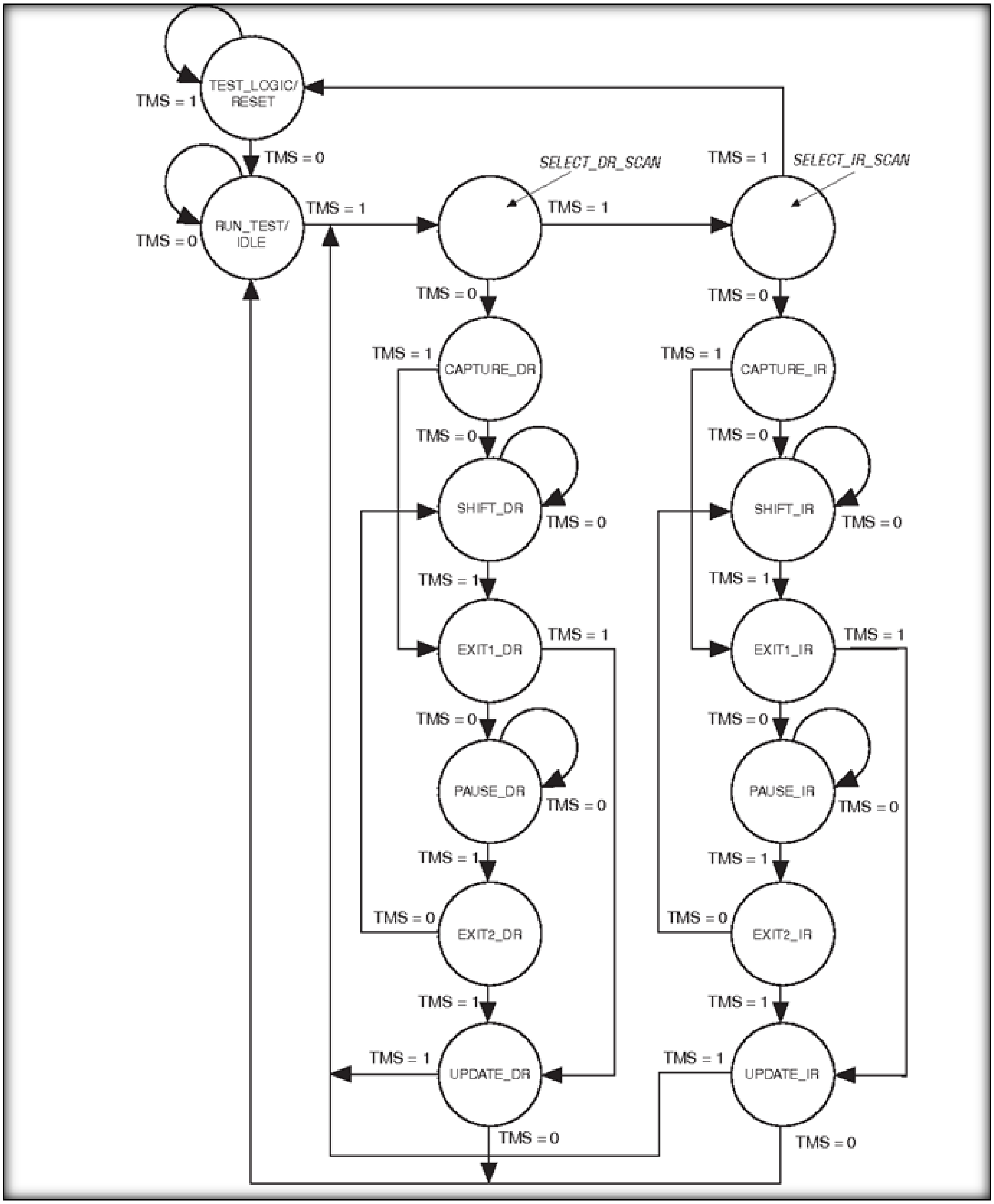

Figure 2-3 : JTAG State Machine [27]

Originally intended of JTAG was to debug and check serviceability of high pin packages of integrated circuits by performing boundary scan. Later on, JTAG was adopted as a multi-purpose debugging and configuration interface. Configuration commands that are sent through the JTAG interface must be enveloped in a specialized JTAG command headers and footers to sequence through the JTAG state machine protocol (refer Figure 2-3), including access to the various JTAG registers.

One of the main advantages of this interface is that the JTAG hardware ports are already built into almost all Xilinx FPGA boards, so using JTAG only requires the use of those four pins.

JTAG is also the highest priority configuration interface, and will always be able to access the configuration module regardless of the configuration mode pins. This interface can program multiple devices, but must do so in a serial chain, (not in parallel like Select-MAP). This interface is typically accessed from an external source, with clock rates up to 66 MHz [16].

The problem with this interface is that only some of its structure is defined in IEEE and rest of the protocol is vendor-specific and its detail are not available for general public usage. However there are commercial-off-the shelf solutions that claim to provide generic JTAG manipulation tool which anybody can use to inject fault in the system. One of the example is JTAGulator [33].

JTAGulator, an open-source hardware tool; can aid in identifying on chip debugging connections from test points, via's, or component pads on a target device which is utilized by researchers, engineers, hobbyists and hackers to retrieve program code or data, modify or alter partial/full contents of a memory, or disturb device operation on-the-fly.

JTAGulator is a microcontroller based solution that can be operated using serial interface. Being a microcontroller based solution JTAGulator has its disadvantages such as its data registers are 32-bit wide and can only support 32-bit JTAG instructions. 7-Series FPGA has instruction length of 6-bits which allows $2^6$ instructions. However that data register for 7-series FPGA can go as much as 256-bits (FPGA bit-stream encryption key length) which cannot be tested using JTAGulator.

### 2.3.2 Select-MAP

Select-MAP is a parallel, high-bandwidth interface with a bi-directional data bus supporting data widths of 8, 16, or 32 bits [18]. SMAP interface provides the ability to configure multiple FPGAs in parallel and can be used with high-speed clock rates as much as up to 100 MHz. The main disadvantage of this interface is that a number of I/O pins (equal to the data bus length) must be reserved during configuration by Select-MAP, and thus are temporarily not available to the user design.

Select-MAP is an FPGA configuration mode that allows user to program as much as up to 3 Xilinx devices in parallel all the while providing simultaneous reading and writing capability through byte-wide ports. All of these devices are however programmed one at a time which is realized through assertion of the correct CS pin at specific time intervals. An external clock source which can be anything such as a download cable, a microprocessor or any other FPGA, is required for successful programming. The data is loaded one byte per CLK pulse.

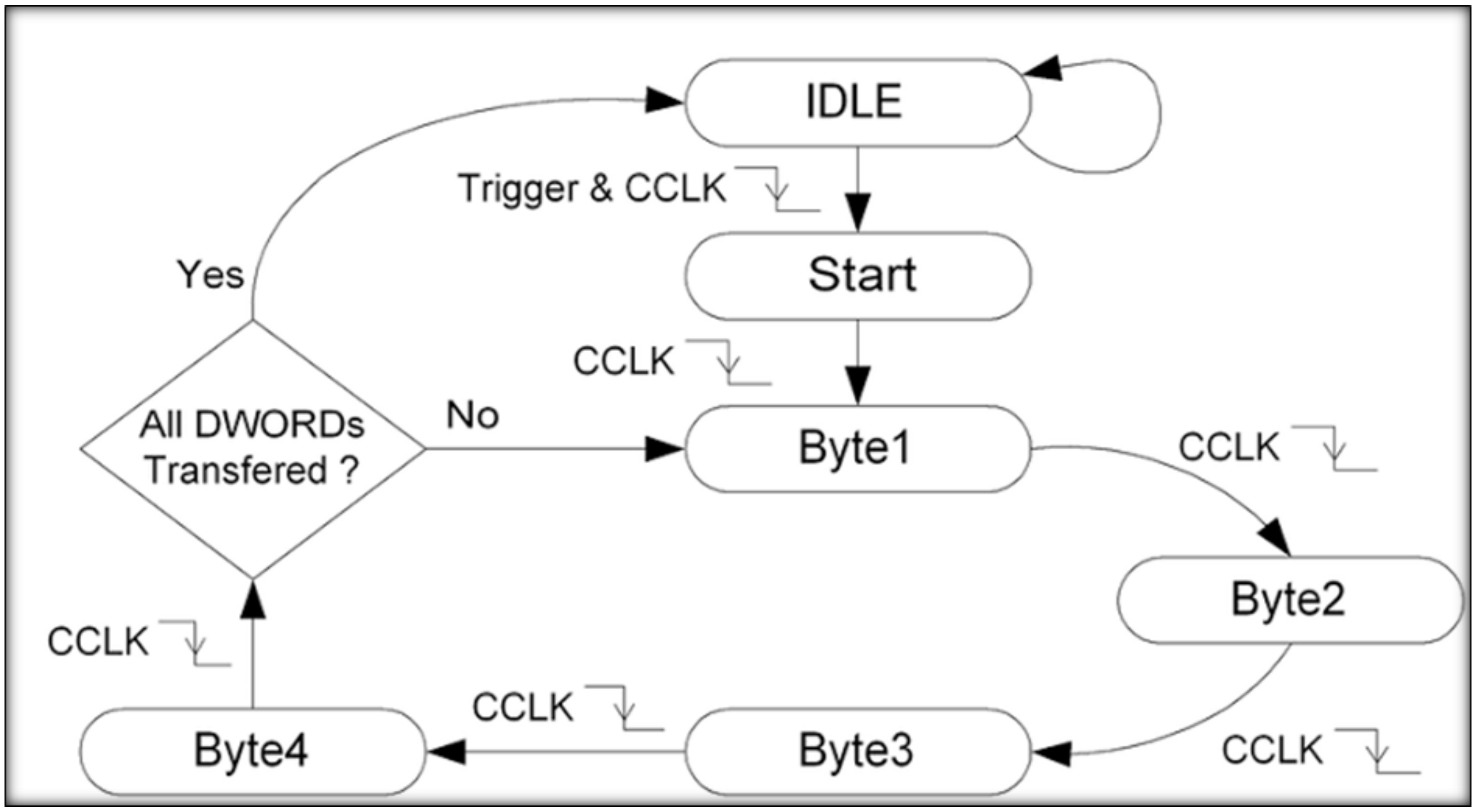

Figure 2-4 : Select-Map Interface FSM [27]

This interface was not explored as an option in this research as this mode is supported by obsolete devices such as Spartan and Virtex families where it is typically used as a primary mode of configuration especially when configuration time is a significant concern or aspect.

### 2.3.3 ICAP

Another option for performing configuration operations is to program the device from within the FPGA itself. Internal configuration Access Port or ICAP can only be accessed from a user design via primitive instantiation. The ICAP allows a user design to reconfigure the device (either partially or fully), after the initial configuration has been completed. However, It is not able to perform the initial configuration.

The ICAP is primarily used for partial reconfiguration. However, ICAP is an important consideration that can also help in development of tamper resistant systems [11]. ICAP can be used to issue IPROG command that can help reset the FPGA to its initial stage in an event of breach. IPROG is a specialized FPGA command that can be sent through the ICAP interface which results in clearing of the entire FPGA configuration memory such as contents of flip-flop, and key expansion memory. It should be noted here that IPROG commands does not the clear the decryption key itself as bit-stream decryption key is stored in BBRAM or E-Fuse. It functionality is almost equivalent to the insertion/applying of the external PROGRAM_B pin. This effectively clears FPGA's configuration memory (flip-flop state, block-RAMs, and configuration data) and can be joined with the KEYCLEARB signal as a reaction to tampering event [11]. However, it is to be noted here that to IPROG command can only be sent to the configuration engine, if the ICAP primitive is present in the user design and the proper order of commands are followed to write to it.

Like Select MAP, one of ICAP primary advantages is its configuration speed, an operating frequency of 100 MHz [21]. The ICAP is the interface of choice for several internal scrubbers in academia and industry. However, due to insertion of ICAP in our IDF design, PL placement and routing was getting disturbed. It failed to place and route essentials signals while conforming to IDF rules so this option was rejected.

However, it is worth mentioning here that ICAP can be side by side in an IDF based design through following proper guideline and careful floor-planning of design.

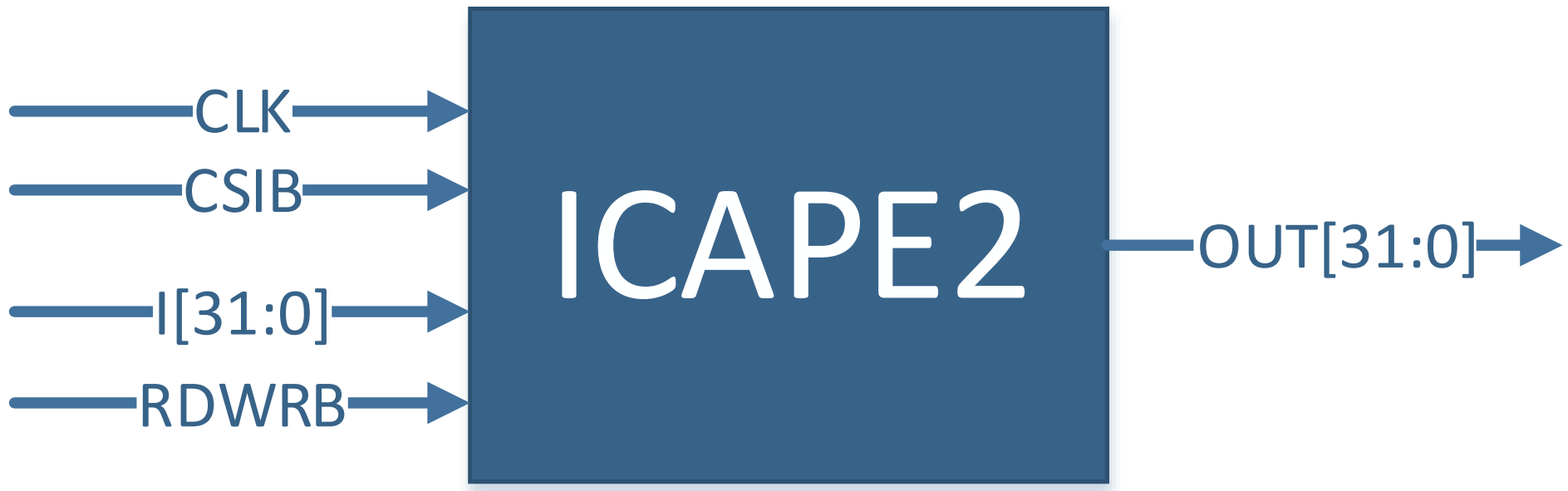

Figure 2-5 : ICAP Primitive for 7-Series FPGA

### 2.3.4 PCAP

The Processor Configuration Access Port or PCAP is a unique interface that enables access from a hard processor to the configuration module. The PCAP is only found on the Zynq-7000 family. It is the bridge between the Zynq's dual-core ARM processing system and the FPGA configuration memory. One of its most important features is that it allows software programs running on the processors to access the configuration module at runtime via configuration commands written in software.

The PCAP clock can run at frequencies as high as 500 MHz, though it usually runs at no higher than 100 MHz for most applications. PCAP clock is also used to clock the bit stream data-path to the PL configuration module. This clock is basically a divided clock; typically PCAP_2x clock. (The supportable clock frequency range for the PCAP clock is can be seen from Zynq TRM [5]). Hence if the user wishes to get a 100 MHz PCAP clock in is design, he must set the PCAP_2X clock bit to 200 megahertz.

PCAP also supports configuration in non-secure mode, which was used in our research as security of the design was not our primary concern. It is worth mentioning here that data transfer rate using PCAP is limited; which is roughly 145 MB/s.

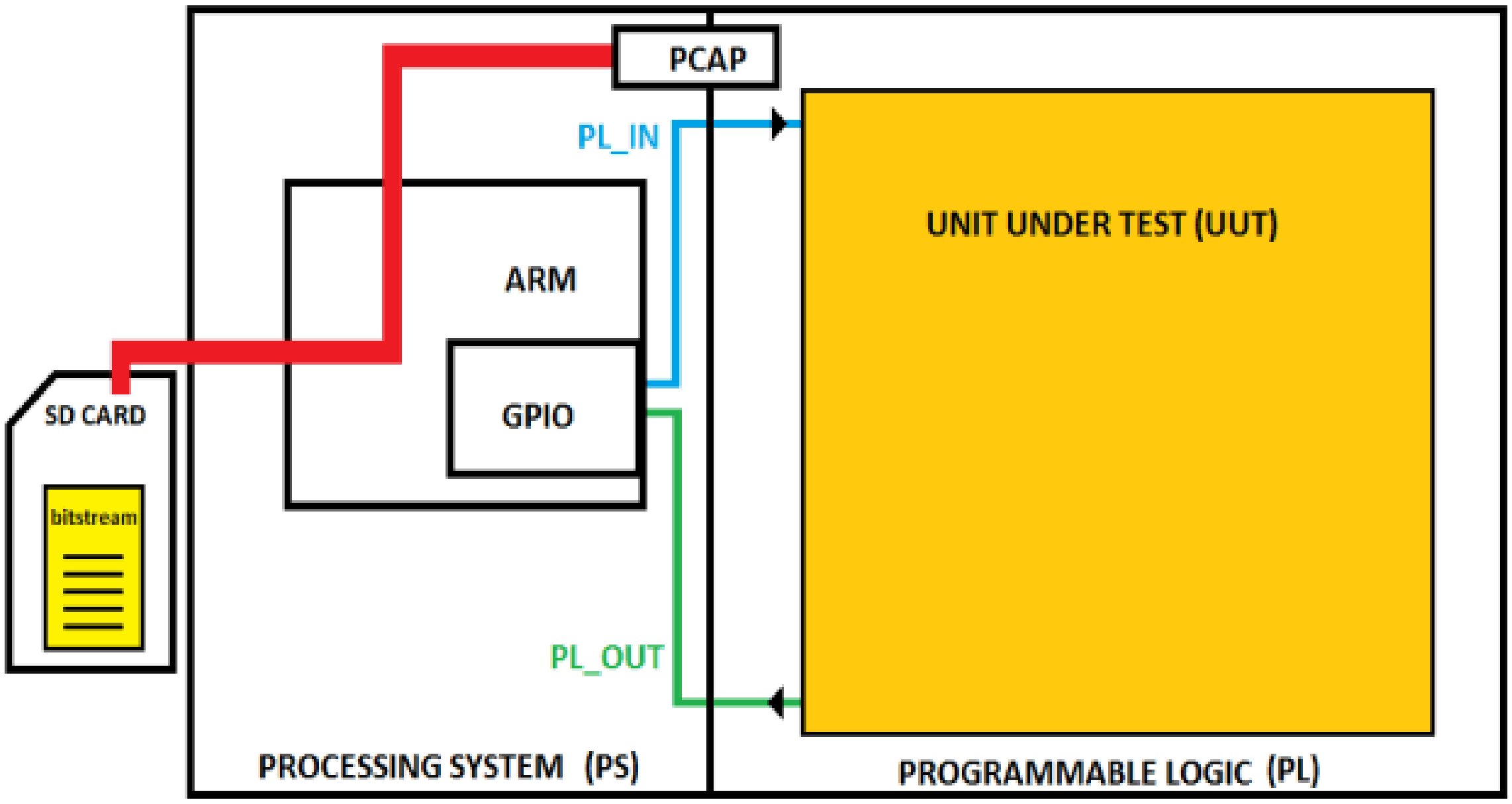

Figure 2-6 : Example of PCAP usage [23]

We know that the PL configuration module although, can handle data rate of 32 bits per PCAP clock cycle, but the overall throughput and performance of PCAP gets limited due to PS AXI interconnect. This approximation is noted from Zynq TRM and assumes a 100 MHz PCAP clock frequency, a 133 MHz APB bus clock frequency, a read issuing capability of four on the PS AXI interconnect, and a DMA burst length of eight [5].

Because the configuration module is the only gateway into the configuration memory, only one of the interfaces may actively process commands through it at any given time. A multiplexer function decides which interface controls the configuration module, so that that interface has exclusive control. The details of this multiplexer function are not publicly documented. A model based on experimental observations and the documentation is given here (refer Figure 2-7), but it does not necessarily represent the actual implementation.

Some configuration interfaces exhibit an “access" priority: if one interface is using the configuration module, a higher priority interface can displace it and gain access even if the original interface is in the middle of a transaction. If a lower priority interface tries to use the configuration port while it is in use by a higher-priority interface, the request will be ignored. JTAG is the highest priority interface, and the read-back CRC is the lowest. It is unknown what the priorities of the other interfaces are. A higher priority interface must issue the DESYNC command to allow a lower-priority interface to gain access.

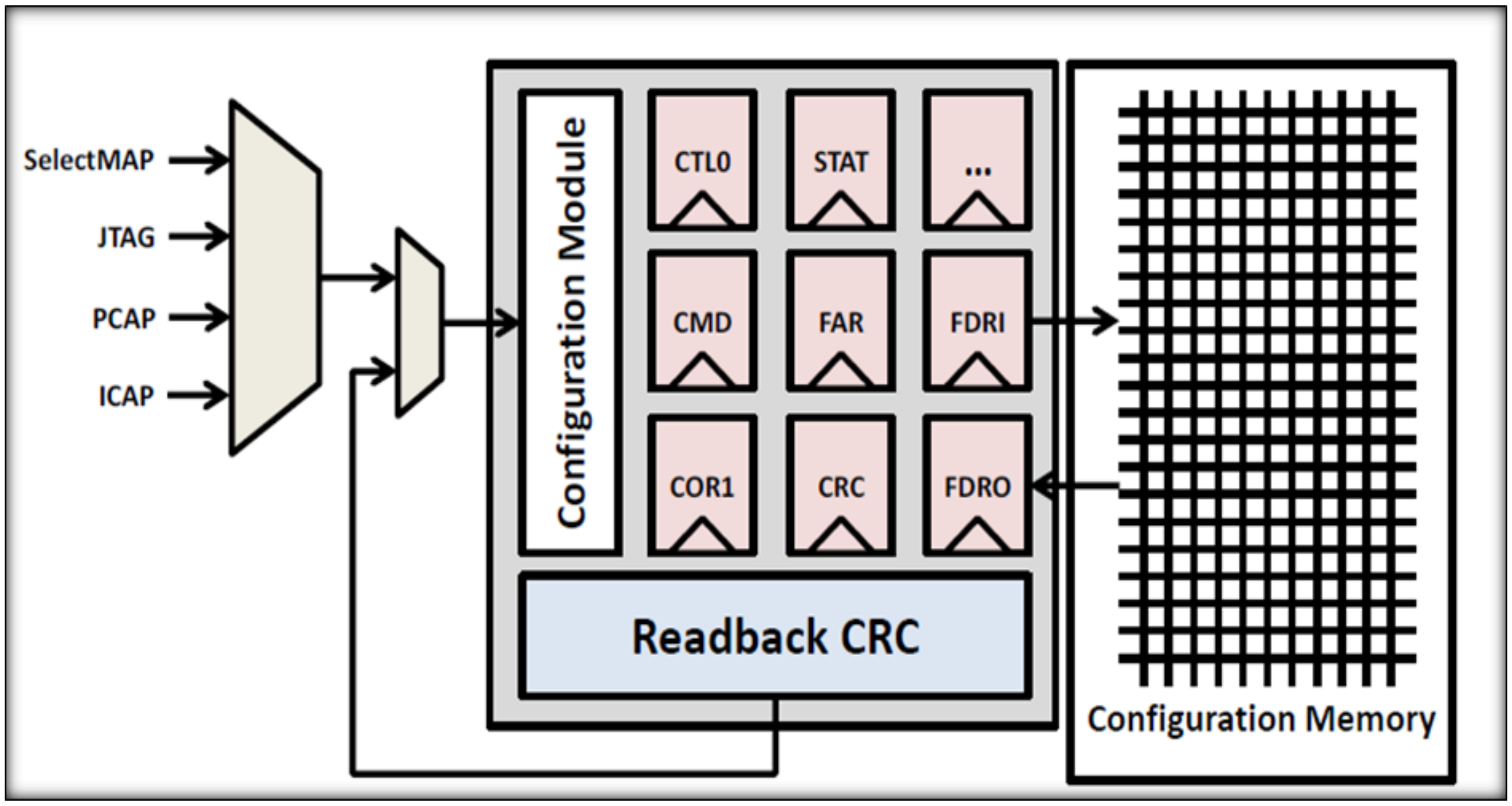

Figure 2-7 : Configuration Interfaces [14]

### 2.3.5 Ionization/Radiation beam

Another method that could be used to inject fault in a system is radiation. FPGAs now-a-days are also being used in high-energy physics experiments at several facilities around the world. These experiments entails the usage of particle accelerators which causes charged particles to travel at high speeds and then collide. Radiation can affect electronics on earth through such means e.g. alpha-particles and neutron-induced Boron Fission. However, all of these radiation sources are expensive and are unavailable in our country. Moreover, this method would only have induced fault. To record and observe the faults we would have had to implement some other method also.

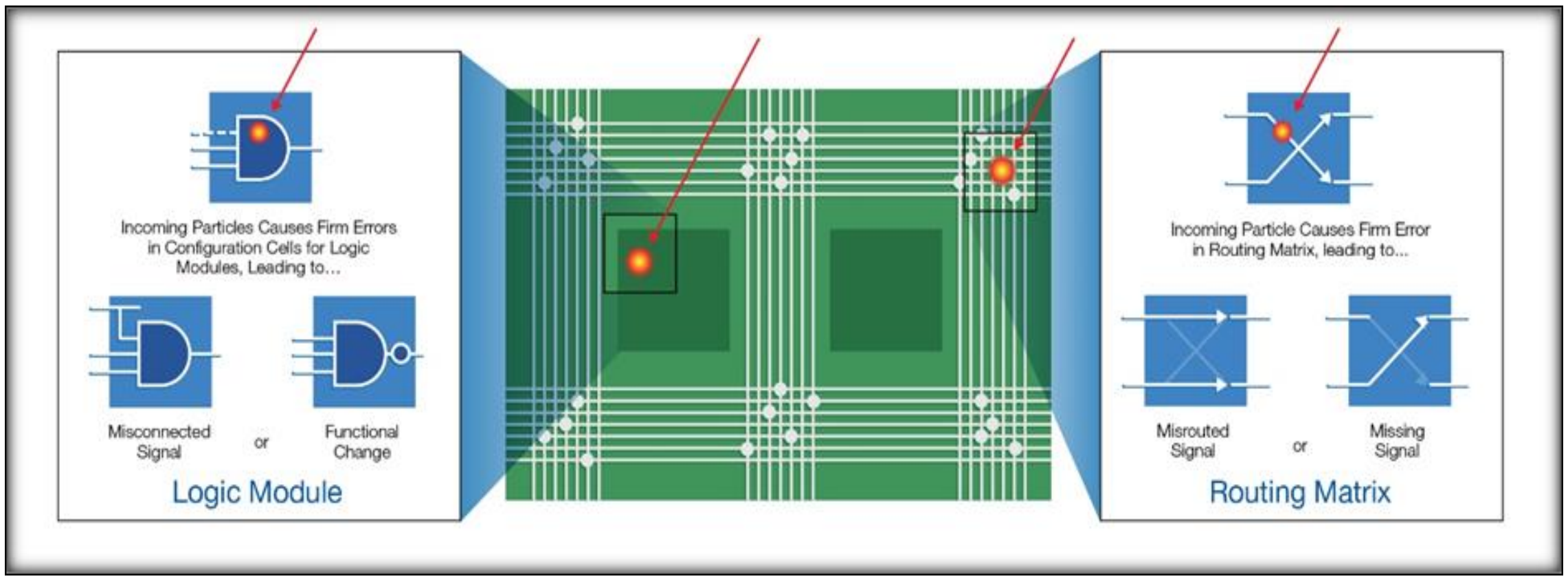

Figure 2-8 : Radiation disrupting SRAM based FPGA Functionality

In our design we decided to utilize PCAP because it gave us two advantages: First, a clear separation between unit under test and fault injector. Secondly, the IDF routing was left undisturbed as we did not have to instantiate any hard macro/ primitives in our design. Using PCAP interface we were able to read any PL frame on demand, modify our desired bits to insert faults and write back to PL using PS which can also help us to keep automated track of faults injected.

## 2.3 Security Aspects of Isolation Design Flow (IDF) and Single Chip Cryptography (SCC)

Another possible question that was briefly explored beside evaluation of IDF was whether or not besides providing fault-tolerance in an SCC environment, could isolation also help in detection and/or prevention of attacks on user's intellectual property. Attacks on Single chip cryptography is a hot topic in the industry and academia. There are methods that if employed correctly could leak user data to unauthorized parties such as Side channel attacks, Power attacks, differential attack etc.

What is the relation between IDF and SCC and whether or not IDF could help increase the modular and IP security was a question that we had initially hoped to pursue as well. However, due to time limitation and unavailability of sophisticated tools and test setup required to thoroughly test, we were not able to pursue it further. Security provided by IDF is an interesting topic that we hope to pursed in the future as well. Our current evaluation however only focuses on performance, area and effectiveness of IDF. A brief comparison on various attacks on SCC was researched whose findings are documented in Table 2-1 for readers interested in pursuing this research area.

Table 2-1 : Attacks on Single Chip Cryptographic Environment

| **Type** | **Paper** | **Contribution** | **Tools Used** | **Evaluation Parameters** |
|---|---|---|---|---|
| Invasive Attacks | Optical Fault Induction Attacks [31] | Provides a mid-range cost invasive attack technique to flip SRAM bits to induce errors. | Wentworth MP-901 Manual Prober, Photoflash lamp, Laser Pointer, PIC16F84 Microcontroller, Serial interface on a PC. | Constant parameters programmable memory with serial interface based read write access to analyze bit flips. |
| | SoC It to EM: Electromagnetic Side-Channel Attacks on a Complex System-on-Chip [15] | Electromagnetic-based leakage on AM335X SOC from execution cryptographic workloads<br>a) Open-SSL server executed AES<br>b) Proprietary AES Co-processor<br>c) NEON core having bit-sliced AES | Beagle Bone Black AM335x SoC, and the NEON core<br>–Tektronix Oscilloscope Model :DPO7104 1GHz,<br>–Signatec digitiser Model :PX14400 400 MS/s,<br>– Langer PA303 pre-amplifier plus various (e.g., low-pass) hardware filters,<br>– Langer RF-3 mini near-field probe set,<br>– Langer ICS105 | 1.Test vector leakage assessment methodology<br>2. Bulk acquire n = 1000 traces, each including = 256 encryption operations<br>3. Deals with systemic noise by searching and filtering for clock scaling and interrupts.<br>4. From each remaining trace, extract a fragment or sub-trace for each |
| | | | IC scanner (or XY-table), – Matlab 2014b (Signal processing toolbox). | encryption Operation; match these with the associated cipher texts. 5. Realign each sub-trace, and discard any corrupted or low-quality cases |
| Non-Invasive Attacks | Temperature Attacks[28] | Data leaking using fan speed measurement. Heat sensing inside FPGA using ring oscillator. Use of temperature bug to monitor and transfer temperature changes to transmit critical information. | Debian Linux on a 2.6.22 (Intel Core2 Duo, 1.80 GHz, and 2 GB RAM PC) MacBook Pro running Mac OS X 10.5.2 (Intel Core2 Duo, 2.16 GHz, and 3 GB of RAM). | PID based control algorithm to stabilize fan speed. Data transfer speed 1bit per minute. along with Ring oscillators to clock a long shift register used to measure regularly the progress of data to sense temperature and frequency |
| | Fault Injection System for SEU Emulation in Zynq SoCs[20] | Analyzes the effect of fault injection on full, partial and One-frame bit-stream using combination of internal and external attack approach | Zed board, PCAP | Inserts error by flipping bits in the bit stream before loading. Mitigates erroneous bit error by re |
| | | | | configuring the system to a known state using golden bit streams from SD card. |
| | Fault Injection, Simple Power Analysis, and Power Glitch Attacks against FPGA-implemented Xoroshiro128+[1] | Performs Single power analysis and glitch attack to make Xoroshiro128+ PRNG lose its randomness by manipulating the initial seed value. | Xoroshiro128+ PRNG, Chip Whisperer-Pro, CW308T XMEGA FPGA | Power consumption of Xoroshiro128+ was measured using Chip Whisperer-Pro.<br>Uses Fault Skipping attack to determine where a particular operation was performed to skips certain operating lines in the source code resulting in the loss of randomness. |
| | Creation and Detection of Hardware Trojans Using Non-Invasive Off-The-Shelf Technologies[9] | 1) Side Channel analysis technique namely power analysis to measure both the power variance, current leakage and traces<br>2) Concentrated heat measurements using an infrared thermometer | Thermal imaging Camera, Basys 3 Artix -7 FPGA, Infinium series 1 GHz Oscilloscope, Digilent Open Scope Oscilloscope ,Infrared Thermometer Gun Powerfix, | A dormant Trojan was implemented in FPGA hardware which when activated caused voltage fluctuation from 3.6V (Normal operation) to 3.9V (Trojan |
| | | 3) Thermal camera test | Flir C2 Heat Camera, Vivado WebPACK Design Suite | operation). Usage of Heat gun and thermal to measure and locate the source (Trojan) of fluctuations in temperature. |

## 2.4 Research Gap

From the literature review it was clear that modular isolation have many useful test cases such as avionics, un-manned probes etc. All of these applications demands that system work flawlessly; without any human intervention. Xilinx provided Isolation solution claims to provide such robustness. However except for Xilinx's self-claim we didn't find any third party verification proof, online which can support Xilinx's claim by providing any concrete evidence. Thus we decided to test ourselves the authenticity of Xilinx claim and level of fault tolerance IDF itself provides to its end users.

# CHAPTER 3 : RESEARCH METHODOLOGY

As discussed previously this thesis is focused on evaluating the Xilinx isolation design flow, specifically Vivado's IDF flow. To this purpose, a design has been implemented on Zynq SoC that conforms to rules and regulation specified by IDF. The verification of all IDF rules have been performed using vivado's isolation verifier to qualify the implemented design. The design was tested on Zynq SoC evaluation board titled "Zed-Board".

The complete design has been implemented in two parts; PL (hardware design under test) and PS section (Fault injection and design evaluation) of Zynq. The design was put to test using SDK where various faults were injected using PS section of Zynq. The complete flow for of the research is described in the subsequent chapters. The thesis is based upon the design that is realized on Zynq SoC specifically Zed-Board. The isolation design flow provided by Xilinx come in two flavors:-

(a) Isolation design Flow for Plan Ahead

(b) Isolation Design flow for Vivado

The rules for IDF defined by both of them do not vary, but the approach for their employment using Vivado tools does. The major rules that are required in applying IDF are stated below for the reader's understanding. However, before the rules of IDF can be stated one should know about the following terms:-

**Ownership** (Logical/physical)—The idea of logical vs. physical ownership is an significant concept to comprehend when making use of IDF.

**Function**—Function can be referred to as a gathering of logic that achieves a specific task or operation (that is, in our case it is AES Encryption Algo).

**Logic**— Digital Circuits that are utilized to realize some function (i.e., flip-flop, LUT's, BRAM, etc.).

**Isolated Region/Pblock**—A physical area on chip die where we can implement and place logic.

**Fence**— A cluster of vacant tiles; in which no logic or routing placement is allowed.

**Trusted Routing**—Trusted routing is enabled; own its own, once the property/attribute HD.ISOLATED is set to “1” or “true” on one or more isolated module. This can be done by either in either GUI or adding the specific constraint in XDC file. These route(s) are a subsection of existing routing resource(s) with the difference being of following restrictions:-[13]

- In the fence region between isolated regions, no entry or exit point exists.
- There is only 1 source and 1 destination region.
- Its entirely exists contained /restricted in the source\destination regions.
- It does not come with-in one fence tile from any other un-intended isolated region.

## 3.1 Vivado’s Isolation Design Flow (IDF)

There are some unique/distinct design rules that must be observed so that FPGA-based Isolation solution is achieved. These rules and regulation that must be considered in design are explained in detail in the subsequent sections. Strict following of these rules is required if one wants to incorporate IDF in design. These rules are as follows:- [13].

## 3.2 IDF Rules

(a) Each function that is required to be isolated must be contained in its own level of hierarchy.

(b) To separate isolated functions, a fence must be used within a single chip.

(c) Input/output buffers must also be placed within isolated modules fences for proper isolation of these buffers. This can only be achieved by user, either instantiating manually or automatically with the help of tools.

(d) If (on-chip) communication between any of the isolated modules or functions is desired it must be realized, through the use of trusted routing. (Vivado PAR tool does this automatically by choosing trusted routes along coincident physical borders).

(e) Top level logic in an isolated designs should be kept to a minimum. In a classic IDF design, only the Clock logic must be present at the top hierarchy.

(f) Top level logic has no placement constraints except than it will not be placed in the fence hence logic in top level can be placed in any isolated region.

(g) Top level logic has no routing restrictions except than it will not violate the fence with used programmable interconnect points or PIPs. Top level routes can be route from, to and through any isolated region.

Once the rules for IDF were thoroughly understood a design was created for the evaluation of IDF. The evaluation design consists of two portions whose descriptions is as follow:-

## 3.2 Hardware Design under Evaluation

The first part of the design was implemented on PL side of Zynq SoC. A reference design was acquired from Xilinx IDF guide for 7-Series FPGA [13]. This reference design contains , two Keccak cryptographic hash redundant modules (ISO_K0 and ISO_K1), whose outputs are routed to another module named comparator block (ISO_Compare), all of these modules are clocked by processor controller (ISO_Controller) module. ISO_Controller is used solely to provide clocks and resets. All of these functions and modules are all isolated within a single FPGA System on chip chip Zynq-7000 device. Visual categorization of these various VHDL sub-blocks used in this design implementation is presented in Figure 3-1.

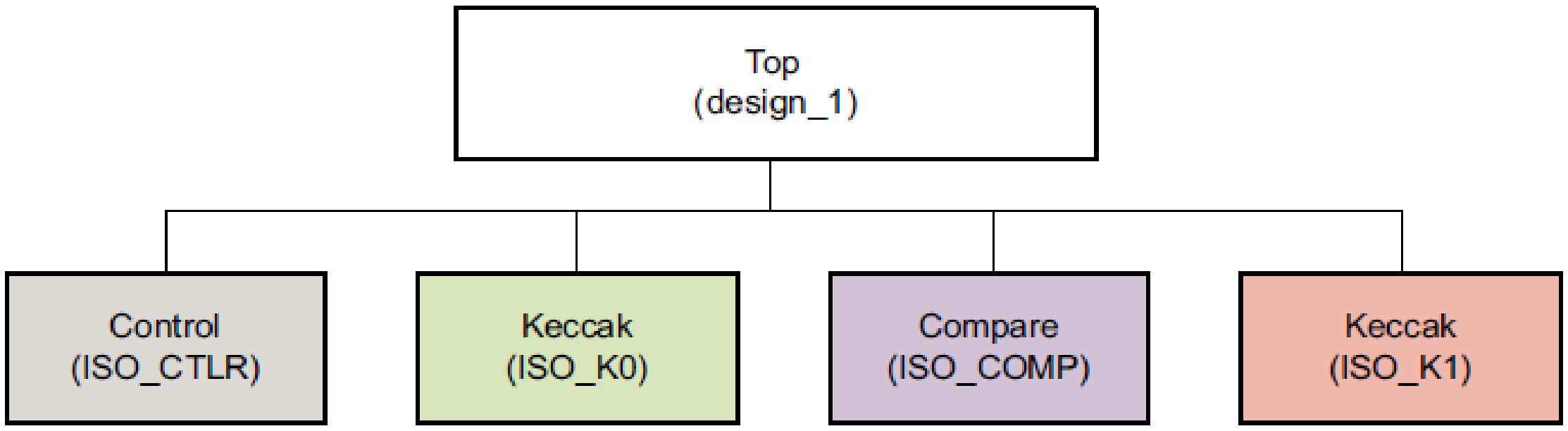

Figure 3-1 : Keccak SCC Design Hierarchy [13]

The floor planning of this design had already been implemented by Xilinx (Figure 3-2). This design has also been implemented on Zed-board.

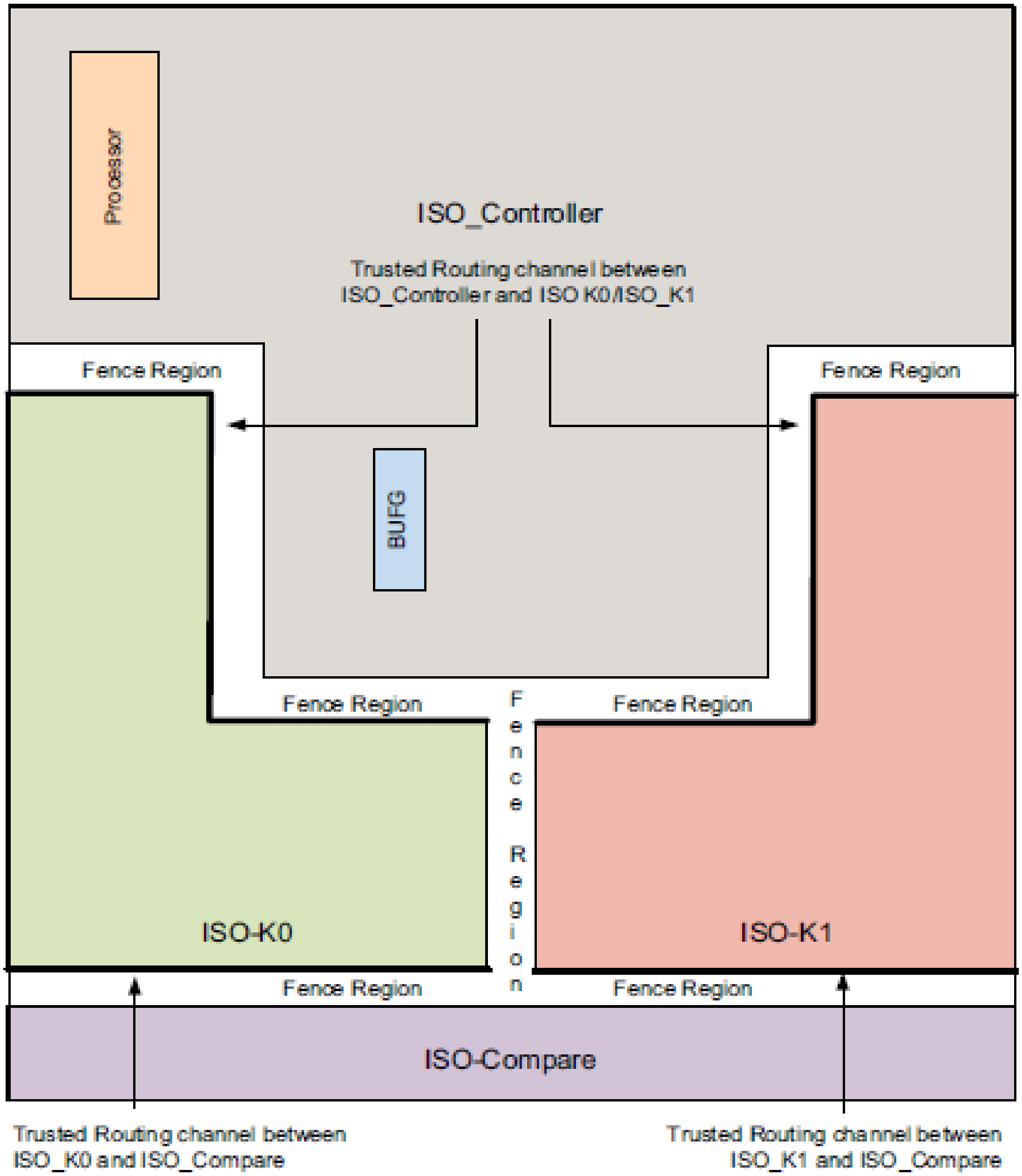

Figure 3-2 : Keccak Floor-plan[13]

This design was taken as reference to implement our own test module. Our design is based on a similar hierarchy with the difference being that the single chip cryptographic function used is AES-256 encryption algorithm. The design is implemented in Verilog language on Vivado 2018.2. The block diagram of our indigenous implementation is as follows:-

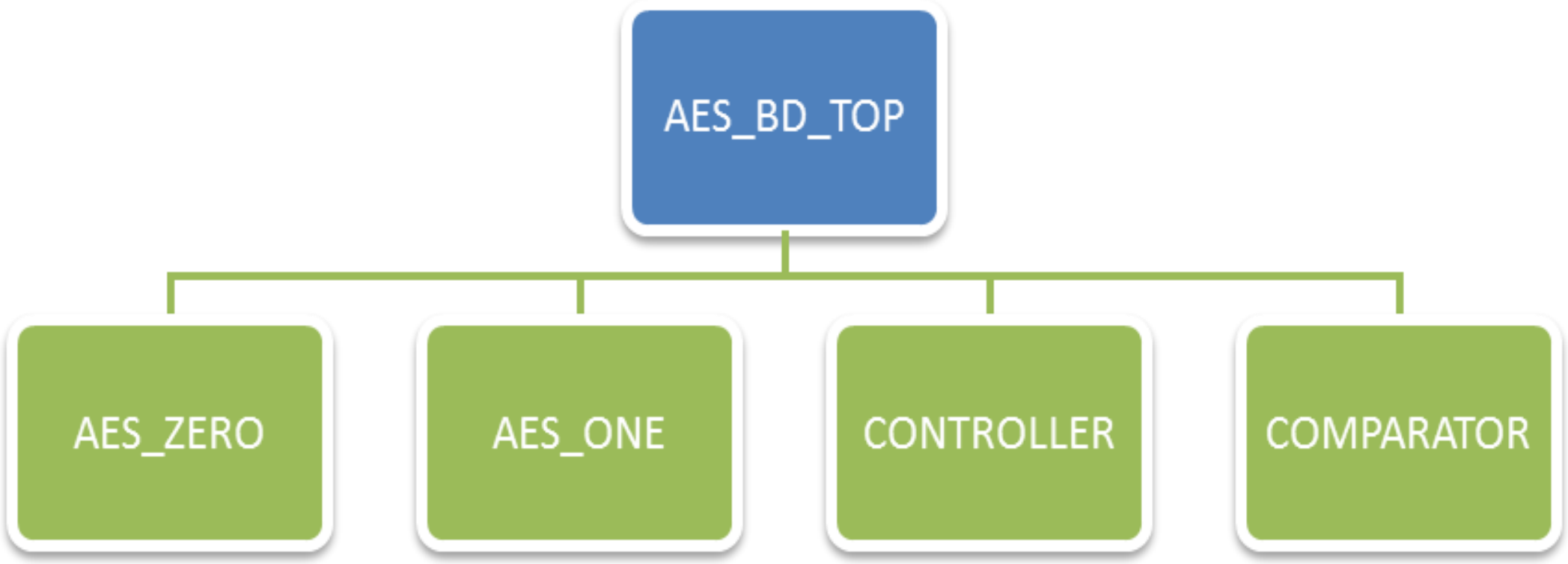

Figure 3-3 : AES-256 Hierarchical Diagram

The design was realized using block diagram feature of Vivado 2018.2. Each module shown in Figure 3-3 lies in its own separate hierarchy and each module is partitioned and mapped to a well-defined location to keep track of all routing and logic track placements. Block design of our implementation is given below:-

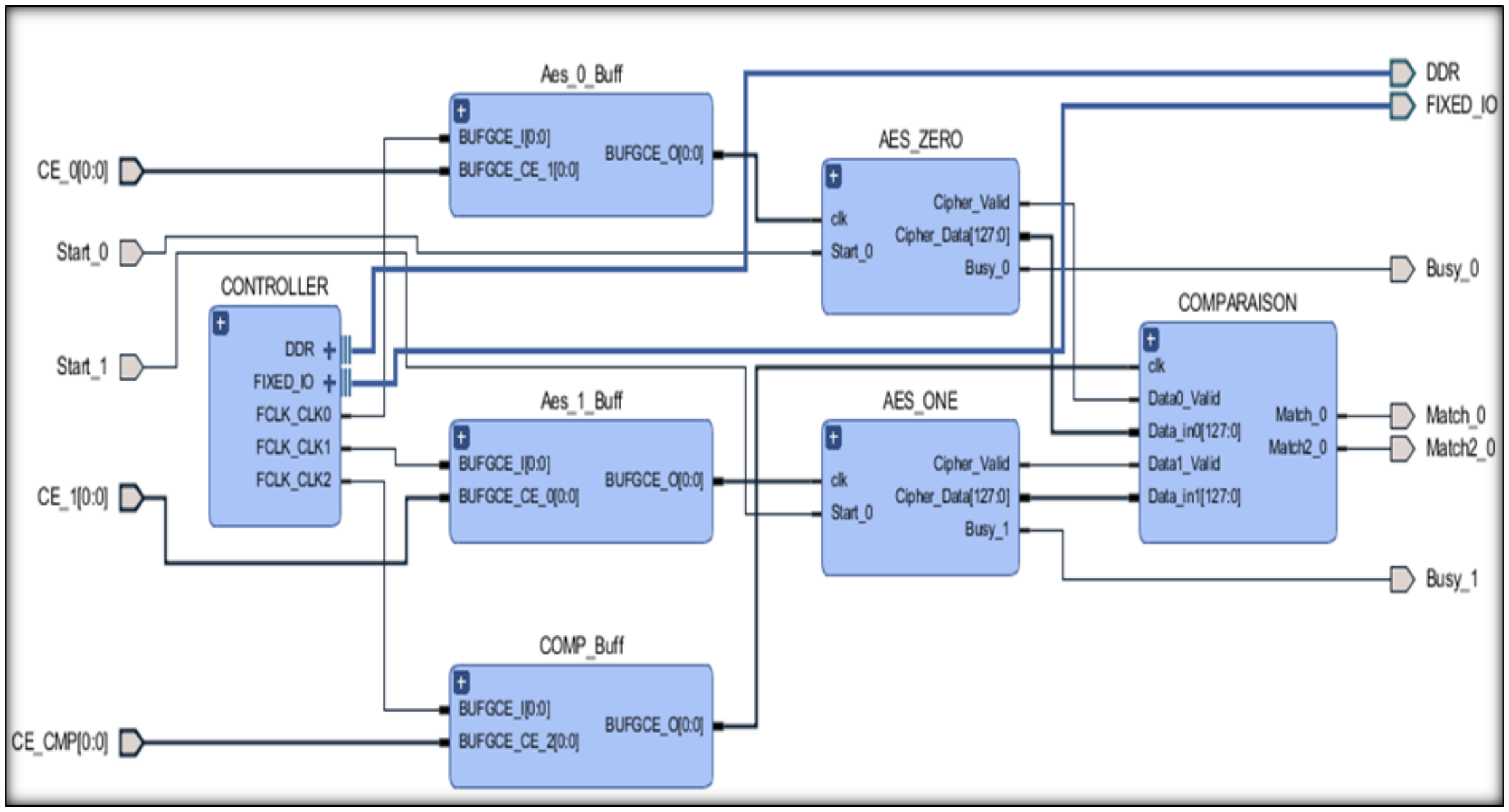

Figure 3-4 : Block Design

AES_ZERO and AES_ONE are two redundant copy of AES implementation. The cipher key used during the process was kept fixed for the sake of simplicity. Each AES module has 4 inputs signals; clock (100Mhz), active low reset signal 4-bit user specified input data that needs to be encrypted and a start signal to initiate the process. Clock and reset signals are provided by Zynq processing system PL- Fabric Clock and Clock Resets. Whereas the clock enable for this modules are controlled and provided externally via PS GPIO's. Rest of the inputs Start_0 and Start_1 are provided through PS GPIO's as well which are controlled programmatically. The output Match is also routed to PS via external wire which is used to monitor and record the output of the modules after each iteration.

When the process of encryption is complete, the output data and a valid data indicator signal is being fed to a comparison module whose purpose is to match both outputs and inform user if the results of both redundant module are same or not using match flag. Match flag is set to one if the outputs matches and vice versa.

Once the purpose and placement of each module was defined, HD.ISOLATED attribute was set on each of them. As mentioned previously this property can be set either in the Vivado Integrated Design Environment (IDE) or the XDC file.An example of the command is:

set_property HD.ISOLATED 1 [get_cells */AES_ZERO_ISO_Wrapper]

Once HD.ISOLATED option is applied on a module/function, all routing & components of that module/function are then isolated. This essentially means, that routing is limited within the boundary of this function, except if it is communicating with another isolated module/function. This also signifies that all components (buffer, logic, PIP's) of that module/function are placed in the corresponding isolated region only. With this, global logic that has been instantiated within an isolated module/function could not be routed globally. The HD. ISOLATED property or attribute needs to be set for each function for the backend implementation tools to use the isolated rules, which allow for floor-planning of an isolated design when using the IDF, but also protects redundant functions from undesired optimization. If the HD.ISOLATED attribute is not set, you can only floor-plan a limited subset of the FPGA components specifically configurable logic block (CLB) slices, block RAM (BRAM), and DSP sites.

To accomplish isolation within a single device, the idea of a fence is presented. The fence is nothing but just a set of unused tiles, as described previously, in which no logic is present. The isolation analysis that were performed by Xilinx claimed that one, or sometimes two, such fencing tiles when present between isolated regions promises that no single point of failure exists in a design that could compromise the isolation between these two regions.

| Horizontal Fence Size | Consequence |
|---|---|
| 1 (minimum required) | All single (1) routes are removed from available routing resources necessary to cross the fence. |
| 2-3 | All single (1) and double (2) routes are removed from available routing resources necessary to cross the fence. |
| 4-5 | All single (1), double (2), and quad (4) routes are removed from available routing resources necessary to cross the fence. |
| 6 or more | A fence this wide cannot be crossed with trusted routes. (Isolation requirements prevent it.) |
| **Vertical Fence Size** | **Consequence** |
| 1 (minimum required) | All single (1) routes are removed from available routing resources necessary to cross the fence. |
| 2-3 | All single (1) and double (2) routes are removed from available routing resources necessary to cross the fence. |
| 4-5 | All single (1), double (2), and quad (4) routes are removed from available routing resources necessary to cross the fence. |
| 6-8 | All single (1), double (2), quad (4), and hex (6) routes are removed from available routing resources necessary to cross the fence. |
| 9 or more | A fence this wide cannot be crossed with trusted routes. (Isolation requirements prevent it.) |

Figure 3-5 : Fence Size Consequences [13]

The minimum fence width can only be determined by a detailed schematic analysis and understanding of user design. Xilinx recommends usage of one tiles fence (though in most cases many more) before the separation between two isolated functions is violated. While it is not compulsory to keep the size of the fence to this minimum (one tile), one can suffer a solid drawback for using broader or larger fences. IDF rules also prevent routing touchdowns in the fence, a fence size of one (minimum allowed) prohibits the use of all routes (to cross the fence) that span only a single tile (singles). Using larger fences may have the consequences listed in Figure 3-5.[13]

Once the floor-plan of the design along with integration of IDF in it was complete, it was necessary to verify that the implemented design adheres to the rule of IDF. For this purpose Xilinx provides Vivado isolation verifier which is available as a separate tcl file that can be integrated into design on-demand. The Vivado Isolation Verifier or VIV is used to verify and ensure that an FPGA design that is partitioned into isolated regions adheres to rigorous standards for fail-safe design. Vivado isolation verifier is a group of design rule checks that aid FPGA developers in fabricating and developing fault-tolerant FPGA applications.

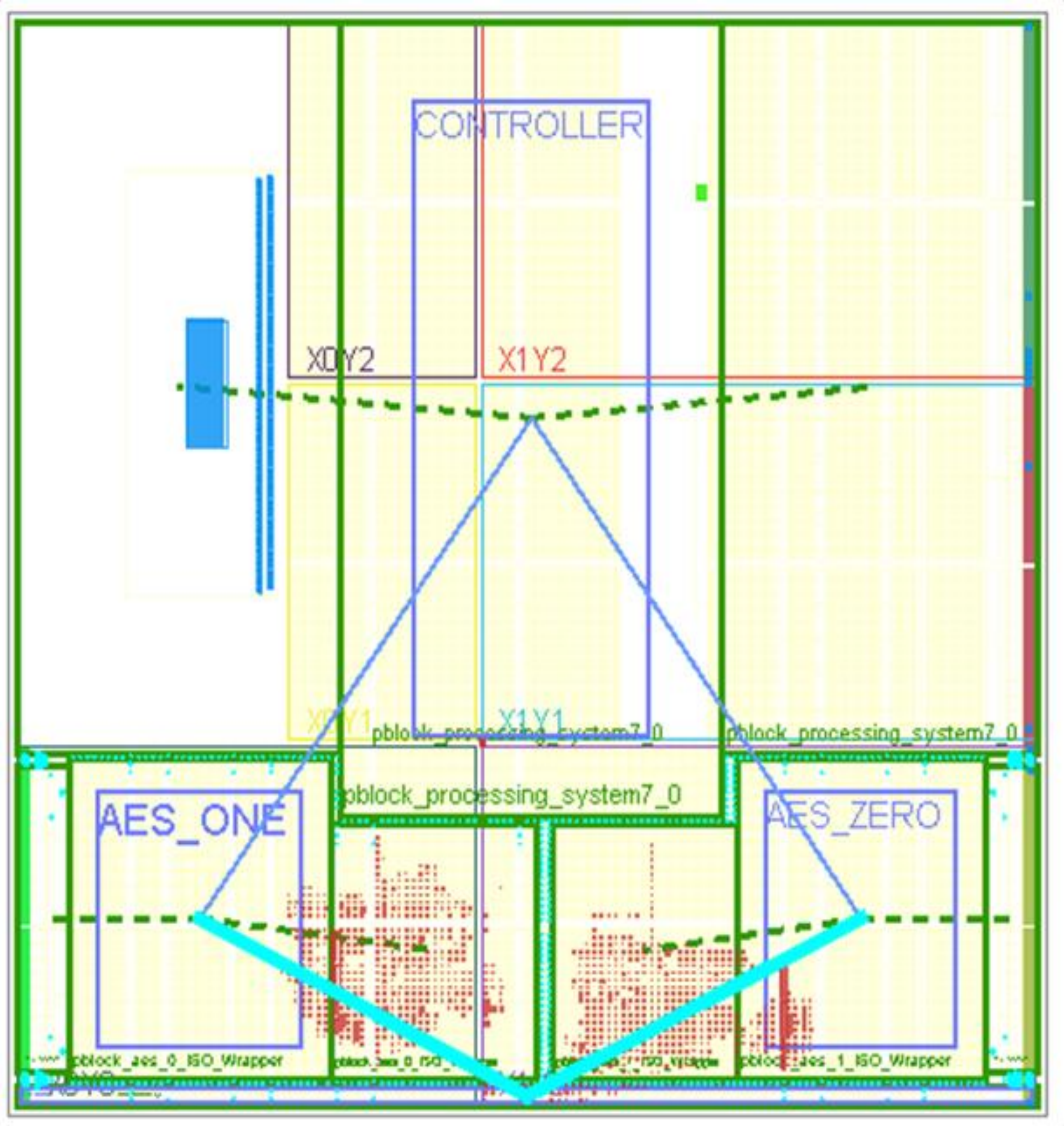

Figure 3-6 : AES Isolated Design Floor Plan

Vivado Isolation Verifier checks the following on the PL floorplan:-

(a) Pins from various isolation groups are not physically neighbors, horizontally or vertically, at the chip die.

(b) Pins from various isolation groups are not physically adjacent at the chip package. Adjacency is defined in eight compass directions: south, north, west, east, northeast, southeast, northwest, and southwest.

(c) Pins from various isolation regions are not co-located in a same IOB bank.

*Note: While VIV considers such conditions as faults, only specific security related applications require this level of bank isolation. In our case as we had to test on a general purpose evaluation board, our source of Data input (DIP switches) lie in the same bank hence this violation occurs when tested using VIV and can be ignored.*

(d) The Pblock constraints placed in the XDC constraints file, are defined such that a minimum of a 1-tile wide fence/seperation exists between isolated regions.

### 3.2.1 Vivado Isolation Verifier

The VIV Tcl script encompasses six design rule checks (DRCs) ; IDF-1 through IDF-6. They are invoked like any other DRCs in the Vivado tool: [13]

1. **IDF-1 Provenance**—IDF-1 documents the circumstances under which a DRC report was generated including tool versions, date, design name, directory, user, platform, and host. No Such violation were found in our design.

2. **IDF-2 I/O Bank violations**—IDF-2 checks all IOBs in banks that have IOBs from numerous isolation groups. No Such violation were found in our design.

3. **IDF-3 Package pin violation**—IDF-3 checks all package pin(s) adjacency violations, should it occur. IDF-3 checks that no two adjacent package pins are from distinct isolation groups. Pin violation exists for our design because it was tested on a general purpose evaluation board who's GPIO's were taken from a single bank (IO Bank-35).

4. **IDF-4 Floorplan violation**—IDF-4 reports all floorplan violations. IDF-4 watches that no isolated region is adjacent to or overlaps another isolated region. No Such violations were found in our design.

5. **IDF-5 Placement violation**—IDF-5 reports all placement violations. IDF-5 checks that no isolated logic or interconnect tile is adjacent to an isolated logic or interconnect tile of a different isolation group. No Such violations were found in our design.

6. **IDF-6 Routing violation**—IDF-6 reports all routing violations and consists of three checks:

   (a) All inter-region nets must have loads in only one isolated region.

   (b) No inter-region net can use nodes that have PIPs in the fence, except clock nets which can have unused PIPs in the fence.

   (c) For any tile containing inter-region nets, all such nets must have a common source and load.

No Such violations were found in our design. With DRC check from VIV being successful, PL side design implementation was complete.

## 3.3 Software Based Design Evaluator

Once the PL side design implementation was complete and successful next step was to implement a fault injection logic. As discussed previously, we chose to use PCAP interface as a primary source of fault injection and data collection. The Zynq SOC contains a new configuration interface known as the Processor Configuration Access Port (PCAP). The PCAP is the gateway for the PS to access the PL and includes a Direct Memory Access (DMA) controller, an AXI bus interface to communicate on the PS AXI interconnect, and a pair of FIFOs (transmit and receive) (refer Figure 3-7). This interface essentially grants the PS easy access to perform configuration operations (like programming a bit-stream) to the PL.

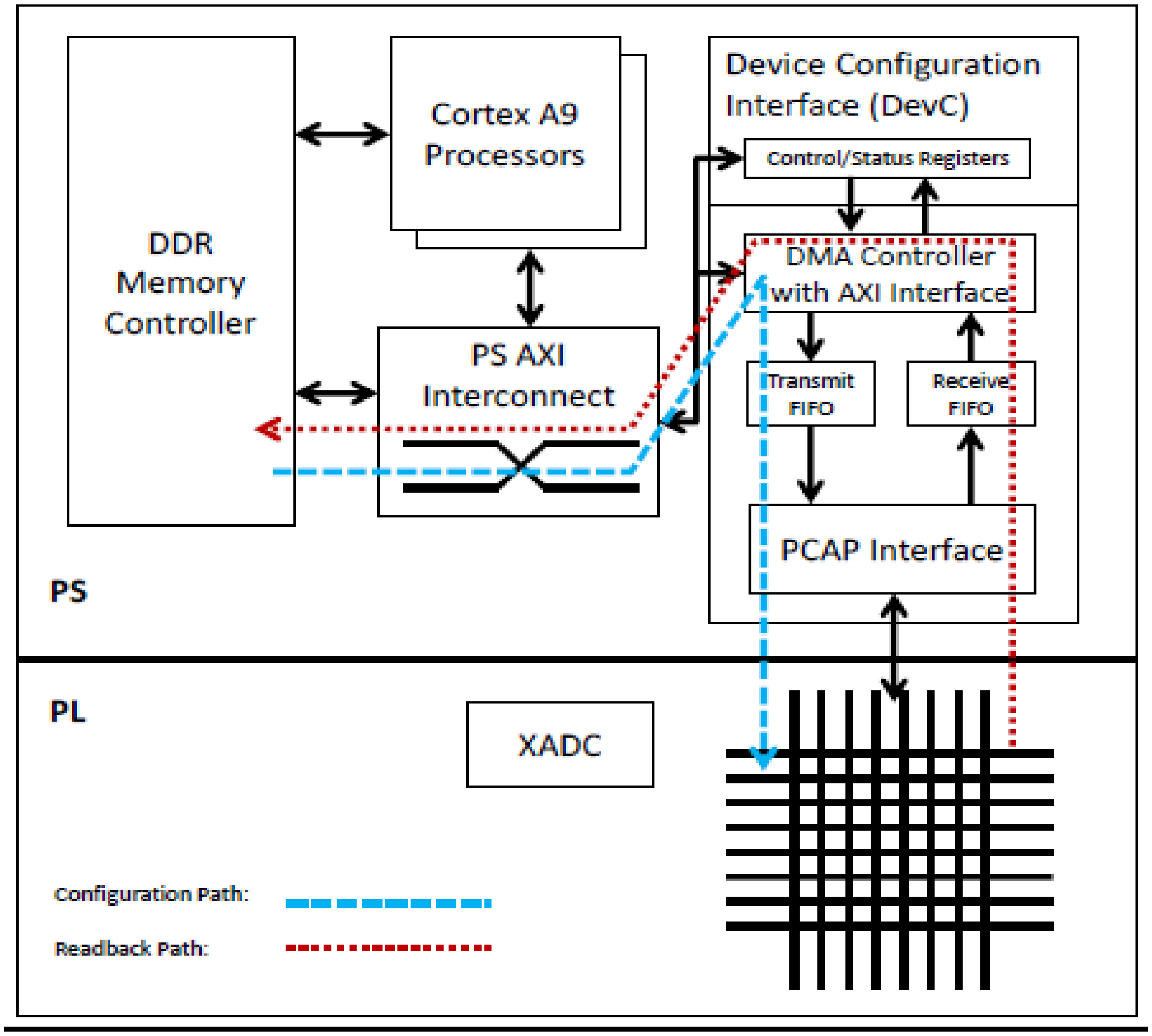

Figure 3-7 : PCAP Interface and Data Transfer Paths [19]

The PCAP is somewhat unique among configuration interfaces in that it does not require a specialized cable or dedicated I/O pins (unlike JTAG or Select-MAP). Instead, the PCAP is accessible to the user through software by using dedicated memory-mapped registers. The PCAP is an internal configuration interface (see Section 5.2), but it is not actually inside the configuration memory. A scrubber using the PCAP does not consume additional logic resources inside the user design like an ICAP scrubber requires. The PCAP scrubber runs on the dual-core ARM processors with the program's memory storage located in DDR or the caches (if enabled).

The PCAP communicates with the PL by transmitting configuration command sequences over Direct Memory Access function which is transferred directly to the configuration module. Configuration module processes these command packets accordingly (see the data path lines in Figure 7.1). The PCAP DMA controller acts as a master on the AXI-bus interface; the controller transfers blocks of data (such as command sequences) from DDR or other large memory location to the transmit FIFO inside the PCAP interface. The PCAP then empties the transmit FIFO and sends the commands one by one to the configuration module. If the configuration commands request a read-back of configuration data, the receiver FIFO will be populated by the PCAP interface receiving the read-back data from the configuration module. The DMA controller then moves the data from the receiver FIFO into a destination memory.

One important note when using the PCAP is that the PCAP must have exclusive configuration module access. With the many interfaces that are available on the Zynq family (Including the ICAP, JTAG, and PCAP), only one can access the configuration logic at a time. For this reason, certain bits in the DevC registers are used to enable the PCAP interface and disable the ICAP interface.

These interfaces also compete with the Read-back CRC (critical to hybrid scrubbing), which has the lowest priority among the configuration interfaces. The Read-back CRC only runs when no other interface is running and when the last interface to access the configuration has properly sent the DESYNC command.

Thus, for a hybrid scrubbing architecture implemented on a Zynq SoC, the PCAP interface would have to release the configuration logic anytime that the Read-back CRC needed to run.

### 3.3.1 Device Configuration Interface (DevC)

There is actually another interface layer between the PCAP and the processors which is called the Device Configuration interface (DevC). The processors communicate to the DevC interface, and the DevC interface forms a bridge with the PCAP interface [55]. PCAP side interfaces directly with the configuration module.

The DevC interface is simply a dedicated address space of accessible memory-mapped registers that allow software to perform configuration operations (refer Figure 3-7 for data-paths).

Configuration scrubbing command sequences are first processed through the DevC interface which are then converted into PCAP protocol and subsequently transferred through the PCAP to the PL [43]. Appendix B gives the sequence for the exact PCAP transfer process. The DevC interface registers facilitate scrubbing in a number of ways:-[5]

(a) Each data transfer from the PS to the PL is initiated by simply writing four registers with the parameters of the transfer in a specific order. ( devcfg.CTRL[PCAP PR] must be set to a `1' to use the PCAP).

(b) A number of status and interrupt bits are dedicated for communicating configuration errors, transfer errors, or internal PL errors. The status registers are valuable in providing debugging information when configuration transfers fail.

(c) Interrupts are also used for reporting non-error conditions such as the DONE status of PCAP and DMA transfers. This allows the PS to know when to resume its program and when to count on the requested PL data being available.

(d) Data transfers require exact widths to be specified. If mismatches occur (i.e. requesting a certain number of read-back frames, but a different number actually gets sent back), the user will know why the transfer failed.

PS can host an OS depending on the OS used and if the caches are enabled, writes to memory mapped registers (including all of the DevC registers used to perform PCAP transfers) will sometimes be cached and will not actually be written through to memory. The caches of the ARM processors are write-back [43]. Hence, the caches must be flushed if they are enabled any time a memory-mapped register is written to. Otherwise, transfers may not actually be initiated at all.

The command sequences and the scrubbing “golden" data are stored in large memories like DDR. The DMA controller can easily transfer these command sequences through the DevC and PCAP interfaces into the PL. Reads from the configuration memory (either read-backs or configuration register reads) are stored in contiguous blocks or software variables in DDR or local CPU registers, respectively. User program can than read and write to the configuration registers and to the configuration memory itself through these DevC/PCAP transfers.

There are some limitations when performing PCAP read-back transfers that must be understood when performing read-back. When a read-back is requested, the returning data comes at a constant rate from the PL whether or not the PCAP receiver FIFO is ready for it. To prevent overflow of the receiver FIFO, the PCAP must transfer this data from the receiver FIFO to the destination memory via DMA over the PS AXI interconnect faster than the configuration module can fill up the receiver FIFO.

The data rate is determined by a combination of the PCAP clock rate and the PS AXI interconnect. When reading back continuously, the DMA controller could hang and freeze the AXI bus if too many frames are being read or if the frames are being read back too fast. Two solutions to handle this issue are presented here.

The first is to read smaller amounts of data (i.e. fewer frames) with long delays (on the order of milliseconds) between each read transfer.

The other solution is to slow the PCAP clock (the default PCAP CLK is 100 MHz). The PCAP CLK can be slowed by writing to a System Level Control Register1 (SLCR) with a larger clock divisor value. We have used this option to slow down read-back (25MHz)

Another limitation of PCAP read-back is that a single read-back request cannot be split over multiple DMA accesses. Sending a command requesting 404 words of read-back data cannot be followed by a read of 202 words, then another read of 202 words. It must read all 404 words in one transfer. The implications of this behavior is that care must be taken when specifying source and destination lengths of the data transfer.

Finally, due to hardware restrictions, all DMA transactions must be organized such that they do not cross a 4 KB boundary. Since read-backs request the number of desired frames plus one dummy frame for the frame buffer per transaction, the most data that the PS can request from the PL in a single transfer are 9 configuration frames + 1 dummy frame = 10 frames. It will be less than 10 frames if the transfer is reading data that crosses a row boundary. As 10 frames equates to 1010 words or 4040 bytes, which is just under the 4096 byte (4 KB) boundary. Attempting to read more than 10 frames in a single transfer will result in DMA transfer errors.

Each read-back performed by the scrubber on the Zynq requires two PCAP transfers:- First is to request the read-back data, and the second is to actually receive the data from the PCAP. While the PCAP has sufficient bandwidth capabilities to write bit streams to the configuration without problems, the read back hardware is smaller i.e. the receiver FIFO is smaller than the transmit FIFO, 1024-Bytes [5] (Refer Figure 3-7) and thus more limited. Timing considerations must be adhered to when implementing a read back on a Zynq device

Description of exact fault injection method performed on PL from PS is necessary before data analysis method is discussed. Upon powering-up of the device a frame template (Refer Appendix-C Frame Template 65) was loaded onto device DRAM. From here user can initiate automatic or manual fault injection onto device.

Automatic mode of fault injection consists of following steps:-

(a) Generate a FAR address using block type, top/bottom, rows, major and minor loop selection.

(b) Read frame data from FAR generated in Step 1 and store its content in DRAM.

(c) Read a word of frame from DRAM, flip a bit and write modified word back to DRAM.

(d) Write the modified frame from DRAM to PL using PCAP.

(e) Run the PL design and capture its response.

(f) Restore the modified Frame to its original state in DRAM.

(g) Write the restored frame from DRAM to PL using PCAP.

(h) Increment bit counter to load next bit and repeat Step 3-6 until the all bits in a frame 101x32-bit words (3232-bits) have been tested.

(i) Go to Step -1 until the complete chip has been tested.

User can also select manual mode of fault injection which consists of following steps:-

(a) Get a FAR from user using serial port and read frame from specified address.

(b) Load the read frame back manually to DRAM using Xilinx Software command line tool command dow -data Frame_Template.bin 0x00200000 where Frame_ Template is the name of the file to load in DRAM at address 0x00200000.

(c) Initiate frame writing function.

(d) Start PL design check function and observe output.

The manual fault injection process is a tedious task which takes approximately 20-22mins execution time to read single frame, modifying a bit, writing it back to PL, record its response and finally restoring of modified bit to its original state and repeating this 3232 times for all frame bits. However manual fault injection can prove useful to quickly test any user specified frame.

# CHAPTER 4 : IDF'S EVALUATION AND ANALYSIS

This chapter describes in detail the exact process that was used to inject fault and record the IDF's behavior. This chapter also explains the test setup used to collect and analyze the data. The findings of data and analysis is also presented in this chapter.

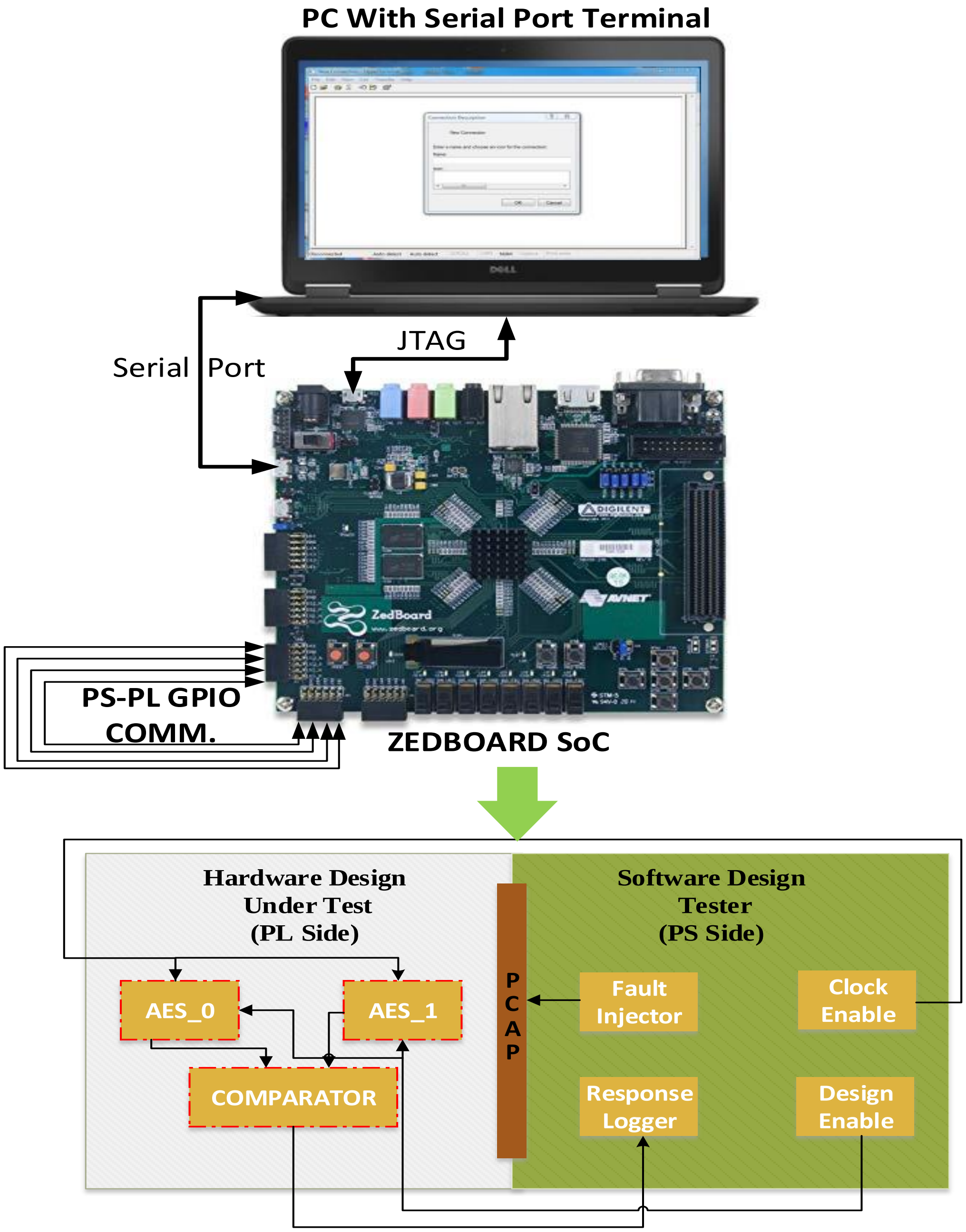

Figure 4-1 : Test Setup

Visual representation of the test setup is shown in above figure for reader's ease of understanding. System is implemented on Zynq SoC; Zedboard in two section; hardware design under test is implemented on PL side while software side evaluation is being performed using PS side.

## 4.1 Evaluation Procedure & Test Setup

The contents of this chapter are structured in following sequence. The initialization process for fault injection, followed by user's selection of automated or manual fault injection. Next the procedure for clock stopping prior to fault injection is discussed, fault injection and design's behavior recording procedure is discussed in the last.

### 4.1.1 Design Initialization

At the power-up of the Zed-board IDF containing design bit file is loaded onto PL followed by fault injection elf file in PS. The PS initialization process is the following. A frame template is loaded into PS DRAM which contains all the necessary header and footer sequence required to configure PL. (Refer Annexure-C).

Followed by this two locations are initialized in DRAM. This locations are 0xFFFF0000 and 0xFFFF0008 which contains the values of error count and error-free counts respectively that will occur in the subsequent process. Once this process is complete the clocks to the modules which are present in the PL are enabled. This modules includes the AES_Zero module which is the primary unit under test in out setup, its redundant golden copy; AES_ONE and the comparison block that compares the output of both AES modules. The clock is controlled by a PS GPIO pin that is connected externally to three PL pins.

These three PL pins are clock enables to utility buffers that are instantiated in the design (Refer Figure 3-4).At this point a menu is displayed to the user using serial port which allows him to make various selections as shown in figure below:-

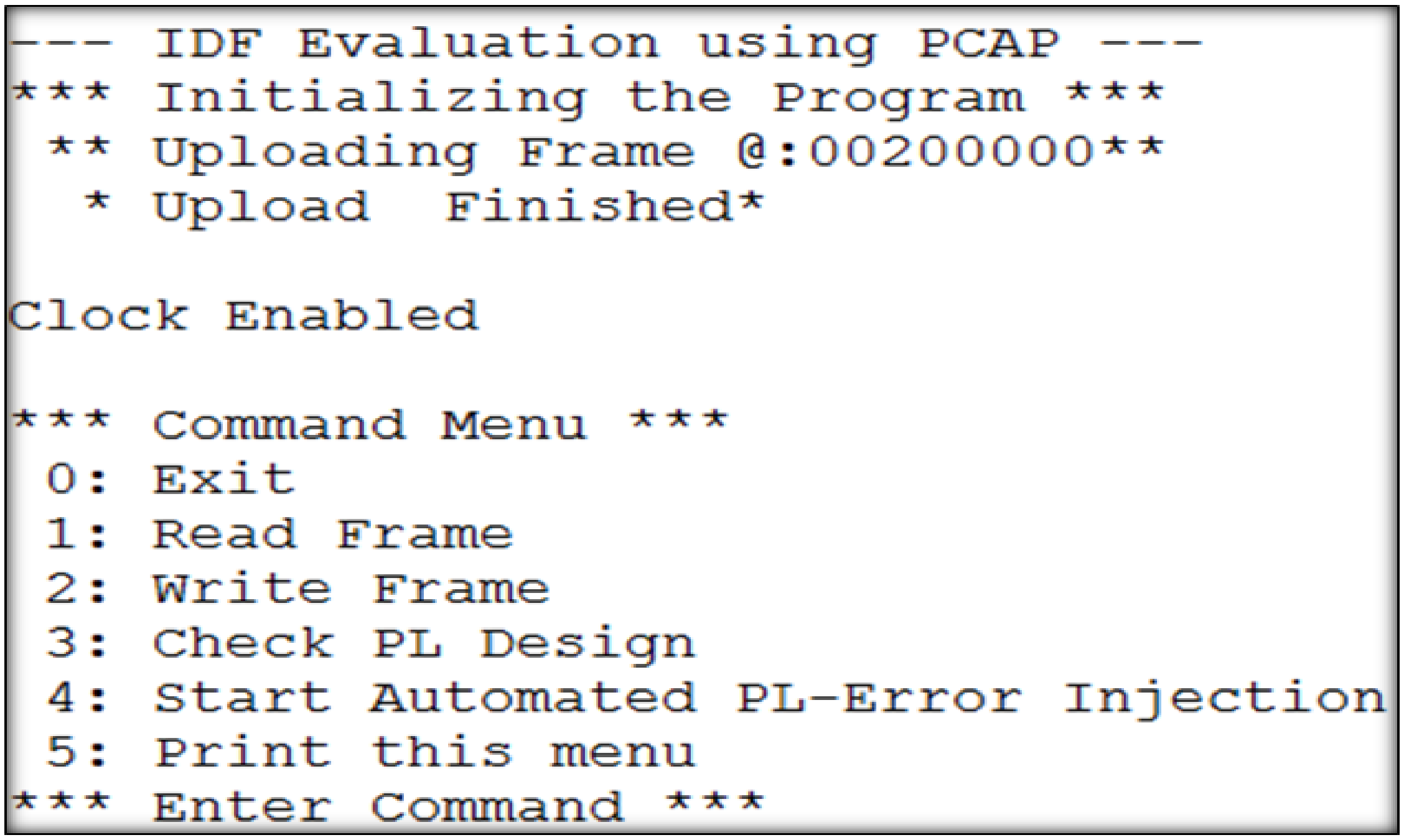

Figure 4-2 : User Menu

### 4.1.2 Fault Injection Methodology

Once the menu is displayed user can initiate either manual or automated fault injection. The manual fault injection take user's input for frame address in which faults are to be injected. The manual fault injection takes approximately 22 minutes to run in which it flips all 3232-bits in a frame are flipped and tested.

The automated fault injection process only takes the process beginning confirmation input from user and then starts generating the frame addresses and testing the entire FPGA. The automated fault injection takes 14-days to test entire FPGA chip. Flipping all 29,600,000 bits, testing the design and restoring it to its original state.

### 4.1.3 Clock Control

Once the user has made a selection to inject fault, the Clocks for the AES_Zero, AES_One and Comparison module is stopped. This is done by pulling down a PS GPIO pin namely PIN 10. This PS GPIO pin is connected externally to three PL inputs which are the clock enables for the aforementioned modules.

### 4.1.4 Error Injection using Bit Flipping

A frame is read back from the PL and entire frame is then written to DRAM. Now the first word of the frame is read back into PS and bit-0 of the frame is xor'ed with a 1 which causes it state to flip from either 0 to 1 or 1 to 0. This modified word is now written back to the DRAM and PS reconfigures the PL using PCAP interface. Once the Frame is written successfully, PS enables the clock by pulling up the clock enable pin mentioned in 0.

### 4.1.5 Hardware Design Verification

After the modified frame is written and clock is restored to PL design it is time to test and record the effect of bit flipping in IDF based design. Start signals are generated to PL via PS GPIO pins 11 and 12. The design under test is running on 50 MHz clock and takes 11 clock cycles to execute completely.

Two further clock cycles are used to register the output of AES_Zero and AES_One and there result comparison. If the result are an exact match the output named Match is set low otherwise it is set to high indicating the error condition. The output Match of PL is also routed back to PS GPIO pin 13. The result of this iteration is recorded in a counter register as well they are written to DRAM reserved locations mentioned in 0.

### 4.1.6 Error Free Design Recovery

Upon recording the behavior of the iteration, the design under test is restored to its original state which is done by repeating previous step which are clock disabling, reading one word from DRAM, XOR'ing bit-0 with 1, writing original word into DRAM and initiating PL frame writing via PS PCAP and enabling the clock to the modules. This process is repeated 32 times for each bit in one word and 101 times for entire frame.

## 4.2 Comparison between IDF and Non-IDF Design

Two designs were made for thorough comparison one having all the IDF and placement constraints and other without IDF constraints but having same placement constraints. The analysis step mentioned in paragraph 4.1 was run for both design and the results were recorded. The results for the procedure are presented in figure below for reader's understanding.

Below figure shows some of the frame address on x-axis where the design under test AES_Zero was placed in Zynq Chip. The y-axis show the errors that was recorded as the result of bit flipping in design. Areas that shows higher peaks of errors in the design is where the routing bits of a frame in design was placed and was highly critical. However once the IDF was enabled and trusted routing was implemented by Vivado, the error rate reduced drastically, by our calculation as much as 12% of error were eliminated.

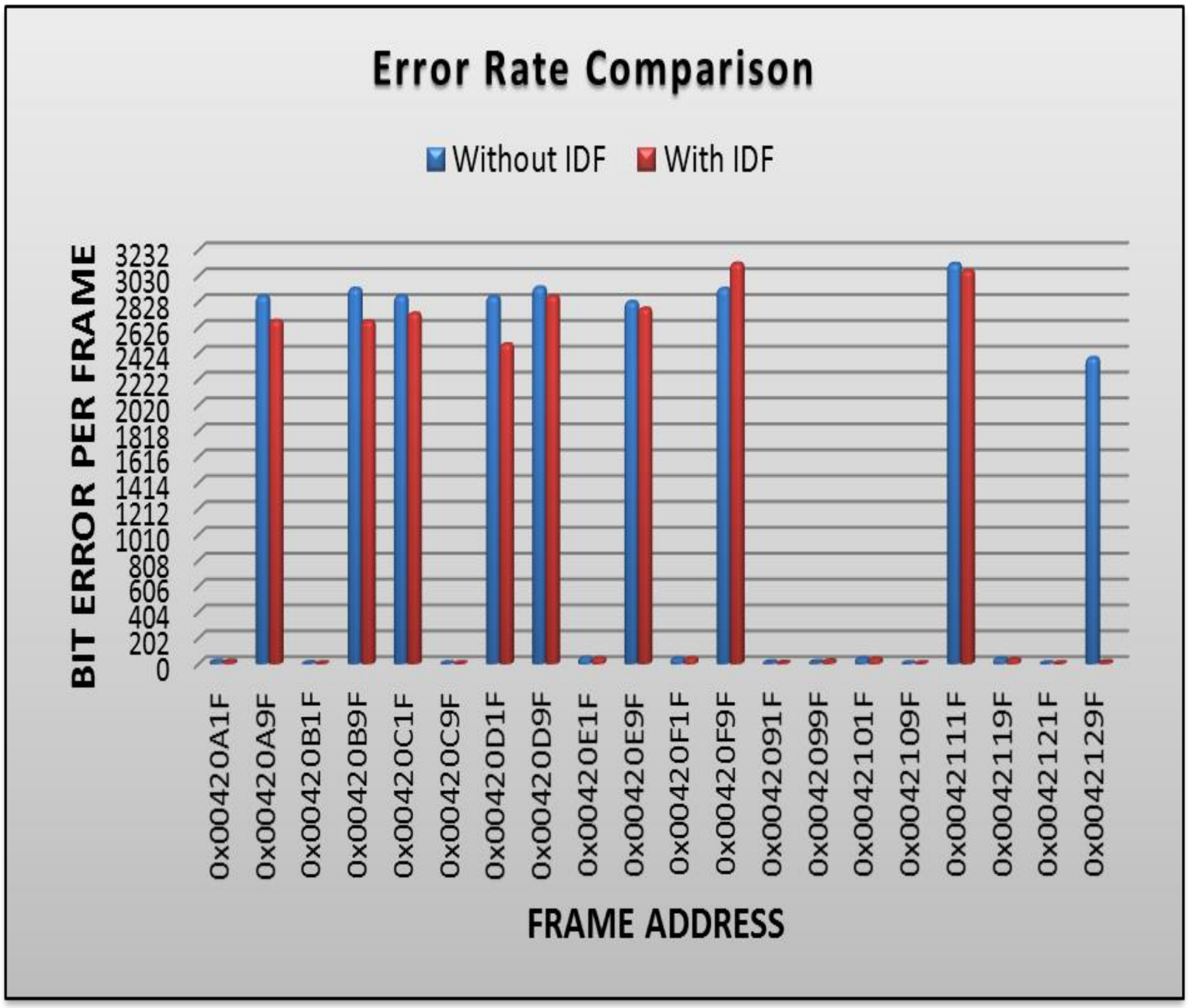

Figure 4-3 : Results Acquired

However the elimination of these error were not free of cost. The trusted routing and design fences that were incorporated in the design reserved resources from the chip and were thus essentially unavailable to our design. Following tables summarizes the total errors injected in the design along with the injections that resulted in system failure errors and those bit that remained dormant and did not affect the design.

Table 4-1 : Fault Injection Summary

| **Injection Type** | **Total Injections** | **Non-critical bits (Injections not resulted in error)** | **Critical bits (Injections resulted in error)** | **Testing Time (min)** |
|---|---|---|---|---|
| Frame Errors (With IDF) | 64640 | 38729 | 25911 | 440 |
| Frame Errors (Without IDF) | 64640 | 37916 | 26724 | 440 |

The details of the resource consumption without IDF incorporated are as follows:-

Table 4-2 : Slice Logic Utilization without IDF

| **Chip Site Type** | **Resources Used** | **Resources Fixed** | **Resources Available** | **Resources Utilization%** |
|---|---|---|---|---|
| Slice Look Up Tables | 2664 | 0 | 53200 | 5.01 |
| LUT as Logic | 2616 | 0 | 53200 | 4.92 |
| LUT as Memory | 48 | 0 | 17400 | 0.28 |
| LUT as Distributed RAM | 48 | 0 | - | - |
| LUT as Shift Register | 0 | 0 | - | - |
| Slice Registers | 1152 | 0 | 106400 | 1.08 |
| Register as Flip Flop | 1152 | 0 | 106400 | 1.08 |
| Register as Latch | 0 | 0 | 106400 | 0.00 |
| F7 Muxes | 536 | 0 | 26600 | 2.02 |
| F8 Muxes | 0 | 0 | 13300 | 0.00 |

Table 4-3 : Slice Logic Distribution without IDF

| **Chip Site Type** | **Resources Used** | **Resources Fixed** | **Resources Available** | **Resources Utilization%** |
|---|---|---|---|---|
| Slice | 836 | 0 | 13300 | 6.29 |
| SLICEL | 621 | 0 | - | - |
| SLICEM | 215 | 0 | - | - |
| LUT as Logic | 2616 | 0 | 53200 | 4.92 |
| using O5 output only | 0 | - | - | - |
| using O6 output only | 2515 | - | - | - |
| using O5 and O6 | 101 | - | - | - |
| LUT as Memory | 48 | 0 | 17400 | 0.28 |
| LUT as Distributed RAM | 48 | 0 | - | - |
| using O5 output only | 0 | - | - | - |
| using O6 output only | 0 | - | - | - |
| using O5 and O6 | 48 | - | - | - |
| LUT as Shift Register | 0 | 0 | - | - |
| LUT Flip Flop Pairs | 423 | 0 | 53200 | 0.80 |
| fully used LUT-FF pairs | 116 | - | - | - |
| LUT-FF pairs with one unused LUT output | 303 | - | - | - |
| LUT-FF pairs with one unused Flip Flop | 294 | - | - | - |
| Unique Control Sets | 30 | - | - | - |

Table 4-4 : Memory Resources Utilization without IDF

| Chip Site Type | Resources Used | Resources Fixed | Resources Available | Resources Utilization% |
|---|---|---|---|---|
| Block RAM Tile | 6 | 0 | 140 | 4.29 |
| RAMB36/FIFO* | 4 | 0 | 140 | 2.86 |
| RAMB36E1 only | 4 | - | - | - |
| RAMB18 | 4 | 0 | 280 | 1.43 |
| RAMB18E1 only | 4 | | - | - |

Table 4-5 : DSP Resources Utilization without IDF

| Chip Site Type | Resources Used | Resources Fixed | Resources Available | Resources Utilization% |
|---|---|---|---|---|
| DSPs | 0 | 0 | 220 | 0.00 |

Table 4-6 : I/O Resources Utilization without IDF

| Chip Site Type | Resources Used | Resources Fixed | Resources Available | Resources Utilization% |
|---|---|---|---|---|
| Bonded IOB | 9 | 9 | 200 | 4.50 |
| IOB Master Pads | 5 | - | - | - |
| IOB Slave Pads | 3 | - | - | - |
| Bonded IPADs | 0 | 0 | 2 | 0.00 |
| Bonded IOPADs | 130 | 130 | 130 | 100.00 |
| PHY_CONTROL | 0 | 0 | 4 | 0.00 |
| PHASER_REF | 0 | 0 | 4 | 0.00 |
| OUT_FIFO | 0 | 0 | 16 | 0.00 |
| IN_FIFO | 0 | 0 | 126 | 0.00 |
| IDELAYCTRL | 0 | 0 | 4 | 0.00 |
| IBUFDS | 0 | 0 | 192 | 0.00 |
| PHASER_OUT/PHASER_OUT_PHY | 0 | 0 | 16 | 0.00 |
| PHASER_IN/PHASER_IN_PHY | 0 | 0 | 16 | 0.00 |
| IDELAYE2/IDELAYE2_FINEDELAY | 0 | 0 | 200 | 0.00 |
| ILOGIC | 0 | 0 | 200 | 0.00 |
| OLOGIC | 0 | 0 | 200 | 0.00 |

Table 4-7 : Clock Resources Utilization without IDF

| **Chip Site Type** | **Resources Used** | **Resources Fixed** | **Resources Available** | **Resources Utilization%** |
|---|---|---|---|---|
| BUFGCTRL | 6 | 0 | 32 | 18.75 |
| BUFIO | 0 | 0 | 16 | 0.00 |
| MMCME2_ADV | 0 | 0 | 4 | 0.00 |
| PLLE2_ADV | 0 | 0 | 4 | 0.00 |
| BUFMRCE | 0 | 0 | 8 | 0.00 |
| BUFHCE | 0 | 0 | 72 | 0.00 |
| BUFR | 0 | 0 | 16 | 0.00 |

The details of the resource consumption after IDF incorporation are as follows:-

Table 4-8 : Slice Logic Utilization with IDF

| **Chip Site Type** | **Resources Used** | **Resources Fixed** | **Resources Available** | **Resources Utilization%** |
|---|---|---|---|---|
| Slice Look Up Tables | 2665 | 0 | 51940 | 5.13 |
| LUT as Logic | 2617 | 0 | 51940 | 5.04 |
| LUT as Memory | 48 | 0 | 17072 | 0.28 |
| LUT as Distributed RAM | 48 | 0 | - | - |
| LUT as Shift Register | 0 | 0 | - | - |
| Slice Registers | 1152 | 0 | 103880 | 1.11 |
| Register as Flip Flop | 1152 | 0 | 103880 | 1.11 |
| Register as Latch | 0 | 0 | 103880 | 0.00 |
| F7 Muxes | 536 | 0 | 25970 | 2.06 |
| F8 Muxes | 0 | 0 | 12985 | 0.00 |

Table 4-9 : Slice Logic Distribution with IDF

| **Chip Site Type** | **Resources Used** | **Resources Fixed** | **Resources Available** | **Resources Utilization%** |
|---|---|---|---|---|
| Slice | 909 | 0 | 12985 | 7.00 |
| SLICEL | 679 | 0 | - | - |
| SLICEM | 230 | 0 | - | - |
| LUT as Logic | 2617 | 0 | 51940 | 5.04 |
| using O5 output only | 0 | - | - | - |
| using O6 output only | 2515 | - | - | - |
| using O5 and O6 | 102 | - | - | - |
| LUT as Memory | 48 | 0 | 17072 | 0.28 |
| LUT as Distributed RAM | 48 | 0 | - | - |
| using O5 output only | 0 | - | - | - |
| using O6 output only | 0 | - | - | - |
| using O5 and O6 | 48 | - | - | - |
| LUT as Shift Register | 0 | 0 | - | - |
| LUT Flip Flop Pairs | 426 | 0 | 51940 | 0.82 |
| fully used LUT-FF pairs | 116 | - | - | - |
| LUT-FF pairs with one unused LUT output | 303 | - | - | - |
| LUT-FF pairs with one unused Flip Flop | 297 | - | - | - |
| Unique Control Sets | 30 | - | - | - |

Table 4-10 : Memory Resources Utilization with IDF

| **Chip Site Type** | **Resources Used** | **Resources Fixed** | **Resources Available** | **Resources Utilization%** |
|---|---|---|---|---|
| Block RAM Tile | 6 | 0 | 126 | 4.76 |
| RAMB36/FIFO* | 4 | 0 | 126 | 3.17 |
| RAMB36E1 only | 4 | - | - | - |
| RAMB18 | 4 | 0 | 280 | 1.43 |
| RAMB18E1 only | 4 | | - | - |

Table 4-11 : DSP Resources Utilization with IDF

| Chip Site Type | Resources Used | Resources Fixed | Resources Available | Resources Utilization% |
|---|---|---|---|---|
| DSPs | 0 | 0 | 200 | 0.00 |

Table 4-12 : I/O Resources Utilization with IDF

| Chip Site Type | Resources Used | Resources Fixed | Resources Available | Resources Utilization% |
|---|---|---|---|---|
| Bonded IOB | 9 | 9 | 192 | 4.69 |
| IOB Master Pads | 5 | - | - | - |
| IOB Slave Pads | 3 | - | - | - |
| Bonded IPADs | 0 | 0 | 2 | 0.00 |
| Bonded IOPADs | 130 | 130 | 130 | 100.00 |
| PHY_CONTROL | 0 | 0 | 4 | 0.00 |
| PHASER_REF | 0 | 0 | 4 | 0.00 |
| OUT_FIFO | 0 | 0 | 12 | 0.00 |
| IN_FIFO | 0 | 0 | 12 | 0.00 |
| IDELAYCTRL | 0 | 0 | 4 | 0.00 |
| IBUFDS | 0 | 0 | 184 | 0.00 |
| PHASER_OUT/PHASER_OUT_PHY | 0 | 0 | 16 | 0.00 |
| PHASER_IN/PHASER_IN_PHY | 0 | 0 | 16 | 0.00 |
| IDELAYE2/IDELAYE2_FINEDELAY | 0 | 0 | 192 | 0.00 |
| ILOGIC | 0 | 0 | 192 | 0.00 |
| OLOGIC | 0 | 0 | 192 | 0.00 |

Table 4-13 : Clock Resources Utilization with IDF

| Chip Site Type | Resources Used | Resources Fixed | Resources Available | Resources Utilization% |
|---|---|---|---|---|
| BUFGCTRL | 6 | 0 | 32 | 18.75 |
| BUFIO | 0 | 0 | 16 | 0.00 |
| MMCME2_ADV | 0 | 0 | 2 | 0.00 |
| PLLE2_ADV | 0 | 0 | 2 | 0.00 |
| BUFMRCE | 0 | 0 | 8 | 0.00 |
| BUFHCE | 0 | 0 | 72 | 0.00 |
| BUFR | 0 | 0 | 16 | 0.00 |

## 4.3 Isolation Design Flow Analysis Report

From the resources utilization obtained in the paragraph 4.2 we have made the following observations from our study. When IDF is enabled in design Vivado synthesizer reserves the resources from the overall design and prohibits there usage in user design thus creating a physical boundary; a separation between each isolated module which restricts the propagation of error in an event of system failure. Following table summarizes the resources comparison of a design with and without IDF.

Table 4-14 IDF Resources Utilization Overhead Summary

| Site Type | IDF Resource Utilization | Percentage Utilization (%) |
|---|---|---|
| Slice LUTs | 1260 | 2.3 |
| LUT as Logic | 1 | 0.01 |
| LUT as Memory | 328 | 1.8 |
| LUT as Distributed RAM | 0 | 0 |
| LUT as Shift Register | 0 | 0 |
| Slice Registers | 2520 | 2.3 |
| Register as Flip Flop | 0 | 0 |
| Register as Latch | 0 | 0 |
| F7 Muxes | 630 | 2.3 |
| F8 Muxes | 315 | 2.3 |
| Block RAMB36/FIFO &RAMB36E1 | 14 | 5 |
| DSP | 20 | 9 |
| Bonded IOB’s | 8 | 4 |
| OUT_FIFO | 4 | 25 |
| IN_FIFO | 4 | 3.17 |
| IBUFDS | 8 | 4.16 |
| OLOGIC | 8 | 4 |
| ILOGIC | 8 | 4 |
| IDELAYE2/IDELAYE2_FINEDELAY | 8 | 4 |
| MMCME2_ADV | 2 | 50 |
| PLLE2_ADV | 2 | 50 |

The size of resources that are marked unusable in the result of IDF thus vary and depends on the following:-

(a) Number of isolated modules.

(b) Height of fence applied (should conform to standards mentioned in Figure 3-5).

(c) Width of fence applied (should conform to standards mentioned in Figure 3-5).

(d) Complexity of logic being implemented in terms of slice and primitive utilization.

(e) Number of global clocking components

(f) Route/signals exempt from isolation

The resource overhead that the IDF thus impacts may vary from design to design however the resource overhead in comparison to its effectiveness against fault containment overweighs itself.

# CHAPTER 5 : CONCLUSIONS & RECOMMENDATIONS

By deep analysis of Xilinx solation design flow we have found that the incorporating IDF in a user design has the several advantages; First, IDF provides logical isolation between various modules in a single-chip module. Second, IDF helps in restriction of error propagation between modules in an event of failure. Third, IDF allows ease of debugging and identification of fault causing module in an event of failure. Fourth, IDF allows reduced probability of failure in case of system failure due to its use of trusting routing and lastly IDF promotes usage of well-defined logical boundary/ separation of module.

Although the advantages that IDF provides are great, we have also observed its following drawbacks, Firstly, IDF rules can be difficult to comprehend for novice user as they sometime require FPGA placement and routing architecture information that normal user may not possess. Secondly, IDF essentially limits users to design in a constraint environment (area wise) so larger design might be difficult to fit onto device along with IDF and lastly, Incorporating IDF in a design, restricts the usage of dynamic partial reconfiguration (side by side) once enabled. If a user wishes to implement IDF and PR side by side he has to do it with the help of third-party tools and API such as Go-Ahead [26] and BITMAN [12] etc.

Xilinx Isolation design flow is a tool that provides its end users a guarantee that the system they are building remains operational in an event of partial or complete system failure. Xilinx achieves this feature by implementing trusted routes between each module implemented in user design on which isolation has been turned on. This accomplishes two things first, that only signals that are deemed as trusted can enter or exit a module and secondly a well-defined boundary between each module is established which restricts Vivado's PAR to route and place with-in an isolated module only, which accomplishes that in case of failure a module does not get affected by other modules only because it was routed through its regions.

Incorporating Xilinx IDF into a solution however comes at a cost of chip resources as the user design gets constraints to specific area, a fence of tiles must be placed around the design which essentially gets subtracted from the total resources available to user. The resources that are thus made unavailable to user as a result of incorporating IDF varies from design to design and those not follow any specific rules or constraints. It's up to the user/designer to constraint the device in such a way that he can achieve the best of area, speed and power all the while achieving his desired functionality.

**Future Work**

During our research we focused on the performance of IDF and the resource overhead that incurs from its incorporation. However, another area that remains to be explored is the security aspect of IDF, and whether or not IDF could be beneficial in providing isolated security to multiple modules present in design by providing logical separation among them. If so, what are the attacks that IDF can prevent or could IDF lead to be an avenue for attacks?

Another challenging task that remains to explore is locating battery-backed RAM location (BBRAM) with-in FPGA (BBRAM is used to securely store key for bit stream encryption) and increasing its security by fencing IDF walls around it.

In our current research we explored the IDF capabilities using DMR which is only useful in detection of an error. For detection as well correction of error we would like to use TMR approach and see how effectively can we utilize the IDF in a TMR based environment.

# ABBREVIATIONS

**AES**: Advanced Encryption Standard

**CLB**: Configurable Logic Blocks

**DMR**: Double Modular Redundancy

**DPA**: Differential Power Analysis

**ECC**: Error Correction Code

**FF**: Flip-Flops

**GSM**: Global Switch Matrix

**HRSoC**: Heterogeneous System On Chip

**IDF**: Isolation Design Flow

**ICAP**: Internal Configuration Access Port

**IPROG**: Internal Programming Interface

**IP**: Intellectual Property

**JTAG**: Joint Task Action Group

**LUT**: Look- Up Table

**P-Block**: Partition Block

**PL**: Processing Logic

**PR**: Partial Reconfiguration

**PS**: Processing System

**PCAP**: Processor Configuration Access Port

**RM**: Reconfigurable Module

**SCA**: Side Channel Analysis

**SCC**: Single Chip Cryptography

**SDF**: Separation Design Flow

**SEU**: Single Event Upsets

**SoC**: System on Chip

**SPOF**: Single Point of Failure

**TMR**: Triple Modular Redundancy

# ANNEXURE-A

## PCAP Transfer Process

With the hardware infrastructure in place, the Zynq SoC can host its own scrubber without requiring a large hardware design to be implemented in its configuration PL or in an external chip that connects via cable to an interface on the Zynq device board. The processors run software code that resides in DDR memory and performs scrubbing operations including read-back, writes to the configuration memory, or accesses to the internal configuration registers in the PL through the DevC and PCAP interfaces

The scrubbing program (.elf file) can be generated by the Xilinx SDK. This .elf file can be programmed into on-board flash memory, loaded onto an SD card from which the Zynq can boot, or be initiated over JTAG using the Xilinx SDK or iMPACT tools.

To perform data transfers through the PCAP and DevC interfaces, the device must first be configured and initialized. The procedure to initialize the PL is as follows:

1. Unlock the DevC interface with the UNLOCK key.

2. Wait for the PL to receive power (PCFG POR B).

3. Wait for the PL to be initialized.

4. Engage the PCAP Bridge by selecting  PCAP for reconfiguration.

5. Make sure the initial PL configuration is done.

After configuring and initializing the PL, configuration command sequences may be sent to the PL using the commands below. Although it may seem redundant to perform some steps repeatedly with multiple command sequences, it is good practice to follow the same sequence of steps to ensure that certain options have not been disabled inadvertently by other applications. The DevC interface greatly simplifies configuration operations down to basic DMA transfers where valid sources, destinations, and lengths are the only operands necessary.

1. Disable the PCAP loopback.

2. Set the data clock rate to be every clock cycle.

3. Queue-up a DMA transfer using the devcfg DMA registers:

(a) Source Address: Location where the new PL bit stream resides.

(b) Destination Address: 0xFFFF FFFF (PL address).

(c) Source Length: Number of 32-bit words in the new PL bit stream.

(d) Destination Length: Total number of 32-bit words in the new PL bit stream. A write to this register last moves the values of all four registers into the command queue.

4. Wait for the DMA and PCAP transfers to finish. When reading from or writing to the configuration logic, a special address code is used to specify the PL or configuration logic as the DMA destination (or source in the case of reading). Writes to the configuration logic differ from reads in that writes only require one DMA transfer, (namely to send the data to the configuration module to be written).

Reads require two transfers: the first to issue the read command, and the second to receive the data sent back by the configuration module.

# ANNEXURE -B

# Configuration Command Sequences for the PCAP

The configuration command sequences given in [18] are specific to the JTAG and Select-MAP on configuration interfaces. While some of the commands in each sequence are required regardless of the interface, less important commands such as NO-OPs and Dummy words may be necessary in variable quantities depending on the configuration interface.

Oftentimes, these less important commands are present merely to flush buffers or give extra time to the configuration logic for processing previous commands. This appendix presents the bare minimum sequences for read and write operations based on experimentation using the PCAP interface. The following sections are divided into write and read operations.

**Write Configuration Data**

Figure 5-1 shows the full command sequence to write to the configuration memory. A dummy frame must be appended after all of the configuration data to flush the frame buffer so that all of the data emerges from the configuration memory.

Writing to the configuration data has a few unique characteristics. It is required to write to the IDCODE register with the actual ID of the FPGA before being able to write to the configuration. In case of Zed-board this ID is 0x23727093. It is also required to write to the CMD register with the write configuration command (WCFG). Typically, it is a wise practice to include a write to the FAR in the sequence with the value of the starting address just in case the FAR changed since the last operation.

| Command Word | Name |
|---|---|
| 0xFFFFFFFF | Dummy |
| 0xAA995566 | Sync Word |
| 0x20000000 | NO-OP |
| 0x30018001 | Write IDCODE |
| 0xXXXXXXXX | Device ID |
| 0x30002001 | Write FAR |
| 0xXXXXXXXX | FRAD |
| 0x30008001 | Write CMD Register |
| 0x00000001 | WCFG |
| 0x30004000 | Write FDRI |
| 0x5XXXXXXX | Data Word Count |
| 0xXXXXXXXX | Data Word 0 |
| ... | Data Words 1 to n-2 |
| 0xXXXXXXXX | Data Word n-1 |
| 0x00000000 | Dummy Word 0 |
| ... | Dummy Word 1 to 99 |
| 0x00000000 | Dummy Word 100 |
| 0x30008001 | Write CMD Register |
| 0x0000000D | DESYNC |

Figure 5-1 : Write Frame Commands

**Read Operations**

The most important difference between the reads and writes with the PCAP is that the write operations are always one transfer while the read operations require two transfers. The DESYNC command is necessarily appended at the end of every write sequence in order to release the configuration logic and allow it to be accessed by the Read-back CRC hardware

Appending the DESYNC command to the end of a PCAP read-back sequence is futile because the data must be read via the second transfer before a DESYNC will have any effect. Sending the DESYNC as a third transfer, after receiving the read-back data in the second Transfer, will correctly be accepted and interpreted to release the configuration module. This explanation is the reason that none of the read operations have DESYNCs at the end of their sequences.

Table 5-1 gives the command sequence to perform a read-back of configuration data. Unlike the write configuration sequence, the read-back command sequence does not require writing to the IDCODE register. Writing to the FAR is optional in this sequence. It is often a good practice to keep the write to the FAR and the read-back operations separate in case there is a need to debug whether the FAR is getting written with the correct value. The data word count must include an extra dummy frame's length in its value to account for the frame buffer otherwise the PL to PCAP transfer will just return a frame of zeroes.

| Step | Port Direction | Configuration Data | Explanation |
|---|---|---|---|
| 1 | Write | FFFFFFFF | Dummy Word |
| | | 000000BB | Bus Width Sync Word |
| | | 11220044 | Bus Width Detect |
| | | FFFFFFFF | Dummy Word |
| | | AA995566 | Sync Word |
| 2 | Write | 02000000 | Type 1 NOOP Word 0 |
| 3 | Write | 30008001 | Type 1 Write 1 Word to CMD |
| | | 0000000B | SHUTDOWN Command |
| | | 02000000 | Type 1 NOOP Word 0 |
| 4 | Write | 30008001 | Type 1 Write 1 Word to CMD |
| | | 00000007 | RCRC Command |
| | | 20000000 | Type 1 NOOP Word 0 |
| 5 | Write | 20000000 | Type 1 NOOP Word 0 |
| | | 20000000 | Type 1 NOOP Word 0 |
| | | 20000000 | Type 1 NOOP Word 0 |
| | | 20000000 | Type 1 NOOP Word 0 |
| | | 20000000 | Type 1 NOOP Word 0 |
| 6 | Write | 30008001 | Type 1 Write 1 Word to CMD |
| | | 00000004 | RCFG Command |
| | | 20000000 | Type 1 NOOP Word 0 |
| 7 | Write | 30002001 | Type 1 Write 1 Word to FAR |
| | | 00000000 | FAR Address = 00000000 |
| 8 | Write | 28006000 | Type 1 Read 0 Words from FDRO |
| | | 482 BA 521 | Type 2 Read 2,860,321 Words from FDRO ( for 7K325T ) |
| 9 | Write | 20000000 | Type 1 NOOP Word 0 |
| | | ... | Type 1 31 More NOOPs Word 0 |
| 10 | Read | 00000000 | Packet Data Read FDRO Word 0 (first 101 words are a dummy frame) |
| | | ... | |
| | | 00000000 | Packet Data Read FDRO Word 2860320 |
| 11 | Write | 20000000 | Type 1 NOOP Word 0 |

Table 5-1 : PL Configuration memory read-back [16]

# ANNEXURE -C

## Frame Template

//Frame Synchronization Header Sequence from UG470.PDF

0xFFFFFFFF, 0xAA995566, 0x20000000, 0x20000000, 0x30008001, 0x00000007, 0x20000000 , 0x20000000, 0x30018001, 0x23727093, /*ZC7020 FPGA ID*/ 0x30002001 , 0x00000000 , 0x30008001, 0x00000001, 0x20000000 ,0x30004000, 0x500000CA, /*CA= 202 Words 1 Fame+1 Dummy*/

// 101- Words Frame Data. Its values can be anything based upon logic implemented

0x00000000, 0x00000000, 0x00000000, 0x00000000, 0x00000000,0x00000000,
0x00000000, 0x00000000, 0x00000000, 0x00000000, 0x00000000,0x00000000,
0x00000000, 0x00000000, 0x00000000, 0x00000000, 0x00000000,0x00000000,
0x00000000, 0x00000000, 0x00000000, 0x00000000, 0x00000000,0x00000000,
0x00000000, 0x00000000, 0x00000000, 0x00000000, 0x00000000,0x00000000,
0x00000000, 0x00000000, 0x00000000, 0x00000000, 0x00000000,0x00000000,
0x00000000, 0x00000000, 0x00000000, 0x00000000, 0x00000000,0x00000000,
0x00000000, 0x00000000, 0x00000000, 0x00000000, 0x00000000,0x00000000,
0x00000000, 0x00000000, 0x00000000, 0x00000000, 0x00000000,0x00000000,
0x00000000, 0x00000000, 0x00000000, 0x00000000, 0x00000000,0x00000000,
0x00000000, 0x00000000, 0x00000000, 0x00000000, 0x00000000,0x00000000,
0x00000000, 0x00000000, 0x00000000, 0x00000000, 0x00000000,0x00000000,
0x00000000, 0x00000000, 0x00000000, 0x00000000, 0x00000000,0x00000000,
0x00000000, 0x00000000, 0x00000000, 0x00000000, 0x00000000,0x00000000,
0x00000000, 0x00000000, 0x00000000, 0x00000000, 0x00000000,0x00000000,
0x00000000, 0x00000000, 0x00000000, 0x00000000, 0x00000000,0x00000000,
0x00000000, 0x00000000, 0x00000000, 0x00000000, 0x00000000,

//Dummy Frame 101 x 32-bits. Its values are fixed to zero's

0x00000000, 0x00000000, 0x00000000, 0x00000000, 0x00000000,0x00000000,
0x00000000, 0x00000000, 0x00000000, 0x00000000, 0x00000000,0x00000000,

0x00000000, 0x00000000, 0x00000000, 0x00000000, 0x00000000,0x00000000,
0x00000000, 0x00000000, 0x00000000, 0x00000000, 0x00000000,0x00000000,
0x00000000, 0x00000000, 0x00000000, 0x00000000, 0x00000000,0x00000000,
0x00000000, 0x00000000, 0x00000000, 0x00000000, 0x00000000,0x00000000,
0x00000000, 0x00000000, 0x00000000, 0x00000000, 0x00000000,0x00000000,
0x00000000, 0x00000000, 0x00000000, 0x00000000, 0x00000000,0x00000000,
0x00000000, 0x00000000, 0x00000000, 0x00000000, 0x00000000,0x00000000,
0x00000000, 0x00000000, 0x00000000, 0x00000000, 0x00000000,0x00000000,
0x00000000, 0x00000000, 0x00000000, 0x00000000, 0x00000000,0x00000000,
0x00000000, 0x00000000, 0x00000000, 0x00000000, 0x00000000,0x00000000,
0x00000000, 0x00000000, 0x00000000, 0x00000000, 0x00000000,0x00000000,
0x00000000, 0x00000000, 0x00000000, 0x00000000, 0x00000000,0x00000000,
0x00000000, 0x00000000, 0x00000000, 0x00000000, 0x00000000,0x00000000,
0x00000000, 0x00000000, 0x00000000, 0x00000000, 0x00000000,0x00000000,
0x00000000, 0x00000000, 0x00000000, 0x00000000, 0x00000000

//De-Synchronization Footer Sequence from UG470.PDF

0x30008001, 0x0000000A, 0x20000000, 0x3000C001, 0x00000100,
0x3000A001, 0x00000000, 0x30008001, 0x00000005, 0x20000000,
0x30008001, 0x0000000D, 0xFFFFFFFF, 0xFFFFFFFF, 0x20000000,
0x20000000